\documentclass[twocolumn,tighten]{aastex6} 

\usepackage{amsmath}

\def\novak{{M31N\,2008-12a}}

\defcitealias{2014A&A...563L...9D}{DWB14}
\defcitealias{2014A&A...563L...8H}{HND14}
\defcitealias{2014ApJ...786...61T}{TBW14}
\defcitealias{2015A&A...580A..45D}{DHS15}
\defcitealias{2015A&A...580A..46H}{HND15}
\defcitealias{2015A&A...582L...8H}{HDK15}
\defcitealias{2016ApJ...833..149D}{DHB16}
\defcitealias{2016ApJ...830...40K}{KSH16}
\defcitealias{0004-637X-847-1-35}{DHG17}

\newcommand{\oonek}{\citetalias{2014A&A...563L...9D}}
\newcommand{\xonek}{\citetalias{2014A&A...563L...8H}}
\newcommand{\ponek}{\citetalias{2014ApJ...786...61T}}
\newcommand{\otwok}{\citetalias{2015A&A...580A..45D}}
\newcommand{\xtwok}{\citetalias{2015A&A...580A..46H}}
\newcommand{\halfk}{\citetalias{2015A&A...582L...8H}}
\newcommand{\othreek}{\citetalias{2016ApJ...833..149D}}

\newcommand{\hstk}{\citetalias{0004-637X-847-1-35}}

\begin{document} 

\received{2017 August 7}
\revised{2017 September 28}
\accepted{2017 September 28}
\slugcomment{Draft version \today, target journal ApJ}

\title{Inflows, Outflows, and a Giant Donor in the Remarkable Recurrent Nova \novak? --- \textit{Hubble Space Telescope} Photometry of the 2015 Eruption}
\shorttitle{\novak: \textit{HST} Photometry of the 2015 Eruption}
\shortauthors{Darnley et al.\ 2017}

\author{
M.~J. Darnley,\altaffilmark{1}
R. Hounsell,\altaffilmark{2,3}
P. Godon,\altaffilmark{4,5}
D.~A. Perley,\altaffilmark{1}
M. Henze,\altaffilmark{6}
N.~P.~M. Kuin,\altaffilmark{7}
B.~F. Williams,\altaffilmark{8}\\
S.~C. Williams,\altaffilmark{1,9}
M.~F. Bode,\altaffilmark{1}
D.~J. Harman,\altaffilmark{1}
K. Hornoch,\altaffilmark{10}
M. Link,\altaffilmark{11}
J.-U. Ness,\altaffilmark{12}
V.~A.~R.~M. Ribeiro,\altaffilmark{13,14,15,16}\\
E.~M. Sion,\altaffilmark{4}
A.~W. Shafter,\altaffilmark{17}
M.~M. Shara\altaffilmark{18}
}

\altaffiltext{1}{Astrophysics Research Institute, Liverpool John Moores University, IC2 Liverpool Science Park, Liverpool, L3 5RF, UK}
\altaffiltext{2}{Department of Astronomy and Astrophysics, University of California, Santa Cruz, CA 95064, USA} 
\altaffiltext{3}{Astronomy Department, University of Illinois at Urbana-Champaign, 1002 W.\ Green Street, Urbana, IL 61801, USA} 
\altaffiltext{4}{Department of Astrophysics \& Planetary Science, Villanova University, 800 Lancaster Avenue, Villanova, PA 19085, USA} 
\altaffiltext{5}{Henry A.\ Rowland Department of Physics \& Astronomy, Johns Hopkins University, Baltimore, MD 21218, USA} 
\altaffiltext{6}{Institut de Ci\`encies de l'Espai (CSIC-IEEC), Campus UAB, C/Can Magrans s/n, 08193 Cerdanyola del Valles, Spain}
\altaffiltext{7}{Mullard Space Science Laboratory, University College London, Holmbury St.\ Mary, Dorking, Surrey RH5 6NT, UK} 
\altaffiltext{8}{Department of Astronomy, Box 351580, University of Washington, Seattle, WA 98195, USA} 
\altaffiltext{9}{Physics Department, Lancaster University, Lancaster, LA1 4YB, UK} 
\altaffiltext{10}{Astronomical Institute, Academy of Sciences, CZ-251 65 Ond\v{r}ejov, Czech Republic} 
\altaffiltext{11}{Space Telescope Science Institute, 3700 San Martin Drive, Baltimore, MD 21218, USA} 
\altaffiltext{12}{XMM-Newton Observatory SOC, European Space Astronomy Centre, Camino Bajo del Castillo s/n, Urb.\ Villafranca del Castillo, 28692 Villanueva de la Ca\~{n}ada, Madrid, Spain} 
\altaffiltext{13}{CIDMA, Departamento de F\'isica, Universidade de Aveiro, Campus de Santiago, 3810-193 Aveiro, Portugal} 
\altaffiltext{14}{Instituto de Telecomunica\c{c}\~oes, Campus de Santiago, 3810-193 Aveiro, Portugal} 
\altaffiltext{15}{Department of Physics and Astronomy, Botswana International University of Science \& Technology, Private Bag 16, Palapye, Botswana}
\altaffiltext{16}{Department of Astrophysics/IMAPP, Radboud University, P.O. Box 9010, 6500 GL Nijmegen, The Netherlands} 
\altaffiltext{17}{Department of Astronomy, San Diego State University, San Diego, CA 92182, USA} 
\altaffiltext{18}{Department of Astrophysics, American Museum of Natural History, 79th Street and Central Park West, New York, NY 10024, USA} 

\begin{abstract}
The recurrent nova M31N\,2008-12a experiences annual eruptions, contains a near-Chandrasekhar mass white dwarf, and has the largest mass accretion rate in any nova system.  In this paper, we present {\it Hubble Space Telescope} ({\it HST}) WFC3/UVIS photometry of the late decline of the 2015 eruption.  We couple these new data with archival {\it HST} observations of the quiescent system and Keck spectroscopy of the 2014 eruption.  The late-time photometry reveals a rapid decline to a minimum luminosity state, before a possible recovery / re-brightening in the run-up to the next eruption.  Comparison with accretion disk models supports the survival of the accretion disk during the eruptions, and uncovers a quiescent disk mass accretion rate of the order of $10^{-6}\,M_\odot\,\mathrm{yr}^{-1}$, which may rise beyond $10^{-5}\,M_\odot\,\mathrm{yr}^{-1}$ during the super-soft source phase -- both of which could be problematic for a number of well-established nova eruption models.  Such large accretion rates, close to the Eddington limit, might be expected to be accompanied by additional mass loss from the disk through a wind and even collimated outflows.  The archival {\it HST} observations, combined with the disk modeling, provide the first constraints on the mass donor; $L_\mathrm{donor}=103^{+12}_{-11}\,L_\odot$, $R_\mathrm{donor}=14.14^{+0.46}_{-0.47}\,R_\odot$, and $T_\mathrm{eff, donor}=4890\pm110$\,K, which may be consistent with an irradiated M\,31 red-clump star.  Such a donor would require a system orbital period $\gtrsim5$\,days. Our updated analysis predicts that the \novak\ WD could reach the Chandrasekhar mass in $<20$\,kyr. 
\end{abstract}


\maketitle

\section{Introduction}

Novae are a sub-class of the cataclysmic variables (CVs), where a white dwarf (WD) accretes hydrogen-rich matter from a donor star within a, typically close, binary system \citep[see][for review articles]{2008clno.book.....B,2014ASPC..490.....W}.  The transferred material usually accumulates in an accretion disk around the WD, but there may also be some element of magnetic accretion at play, depending upon the strength of the WD's magnetic field.  Novae distinguish themselves from CVs by virtue of their typically elevated WD mass accretion rates ($\dot{M}_\mathrm{acc}$) and by the nova eruption itself -- a thermonuclear runaway within the accreted envelope on the WD surface \citep[see][]{1976IAUS...73..155S}.  All novae are inherently recurrent, but their inter-eruption period depends upon the WD mass ($M_\mathrm{WD}$) and $\dot{M}_\mathrm{acc}$.  Systems that combine a large $M_\mathrm{WD}$ with a high $\dot{M}_\mathrm{acc}$ exhibit the shortest recurrence periods, and have often been observed in eruption more than once -- the recurrent novae \citep[RNe; see][]{2010ApJS..187..275S}.  Observed recurrence periods lie in the range $1\leq P_\mathrm{rec}\leq98$\,years \citep[see][respectively]{2014A&A...563L...9D,2009AJ....138.1230P}, where both ends are probably limited by selection effects \citep[see, for e.g.,][]{2014ApJ...793..136K,2016ApJ...819..168H,2017ApJ...834..196S}.

In most cases, the high values of $\dot{M}_\mathrm{acc}$ in the RNe are driven by elevated mass loss rates from evolved donors \citep{2012ApJ...746...61D}.  This is observed to be via Roche lobe overflow of a sub-giant donor (e.g., U\,Scorpii), or by accretion from the stellar wind of a giant \citep[e.g., RS\,Ophiuchi, see e.g.,][]{2008ASPC..401.....E} --  both mechanisms lead to an accretion disk around the WD.  A handful of RNe, possibly `transient' (rather than long-term) recurrents, such as T\,Pyxidis, may show evidence of elevated mass transfer driven by the irradiation of their main sequence donors \citep{2000A&A...364L..75K,2014ApJ...784L..33G}.

\novak, a RN residing within M31, is the most extreme nova system yet discovered.  With an observed $P_\mathrm{rec}\simeq1$\,year, it is the prototype of the newly emerging class of `rapidly recurring nova'; those with $P_\mathrm{rec}\lesssim10$\,years.  First detected in 2008, \novak\ has been discovered in eruption every year since \citep[2008--2016;][]{2014A&A...563L...9D,2015A&A...580A..45D,2016ApJ...833..149D,2016ATel.9848....1I}, with three previous eruptions recovered from archival X-ray observations \citep[1992, 1993, 2001;][]{2014A&A...563L...8H,2014ApJ...786...61T}.  For reference, the observed eruption history is summarised in Table~\ref{eruption_history} \citep[see][for a detailed description]{2016ApJ...833..149D}.  By analysis of the eight eruptions between 2008--2015, \citet[hereafter \othreek]{2016ApJ...833..149D} reported $P_\mathrm{rec}=347\pm10$\,days.  However, when including the earlier X-ray detections, \citet[hereafter \halfk]{2015A&A...582L...8H} suggested that $P_\mathrm{rec}$ could even be as short as $174\pm10$\,days.

\begin{table}
\caption{A summary of the twelve observed eruptions of \novak.\label{eruption_history}}
\begin{center}
\begin{tabular}{lll}
\hline
\hline
Eruption date\tablenotemark{a} & Inter-eruption & References\\
(UT) & time-scale (days)\tablenotemark{b} \\
\hline
(1992 Jan.\ 28) & \nodata & 1, 2 \\
(1993 Jan.\ 03) & 341 & 1, 2 \\
(2001 Aug.\ 27) & \nodata & 2, 3 \\
2008 Dec.\ 25 & \nodata & 4 \\
2009 Dec.\ 02 & 342 & 5 \\
2010 Nov.\ 19 & 352 & 2 \\
2011 Oct.\ 22.5 & 337.5 & 5, 6--8 \\
2012 Oct.\ 18.7 & 362.2 & 8--11 \\
2013 Nov.\ $26.95\pm0.25$ & 403.5 & 5, 8, 11--14 \\
2014 Oct.\ $02.69\pm0.21$ & $309.8\pm0.7$ & 8, 15 \\
2015 Aug.\ $28.28\pm0.12$ & $329.6\pm0.3$ &14, 16--18\\
2016 Dec.\ 12.32 & 471.72 & 19, 20\\
\hline
\end{tabular}
\end{center}
\catcode`\&=12
\tablenotetext{a}{Eruption dates in parentheses have been estimated based on an extrapolation of available X-ray data \protect \citep[see][]{2015A&A...582L...8H}.}
\tablenotetext{b}{The inter-eruption time-scale is only given when consecutive eruptions have been detected (assuming $P_\mathrm{rec}\simeq1$\,year).}
\tablecomments{Compact version of a Table originally published by \citet{2014ApJ...786...61T} and updated by \citet{2016ApJ...833..149D}.}
\tablerefs{(1)~\citet{1995ApJ...445L.125W}, (2)~\citet{2015A&A...582L...8H}, (3)~\citet{2004ApJ...609..735W}, (4)~\citet{2008Nis}, (5)~\citet{2014ApJ...786...61T}, (6)~\citet{2011Kor}, (7)~\citet{2011ATel.3725....1B}, (8)~\citet{2015A&A...580A..45D}, (9)~\citet{2012Nis}, (10)~\citet{2012ATel.4503....1S}, (11)~\citet{2014A&A...563L...8H}, (12)~\citet{2013ATel.5607....1T}, (13)~\citet{2014A&A...563L...9D}, (14)~\citet{2016ApJ...833..149D}, (15)~\citet{2015A&A...580A..46H}, (16)~\citet{2015ATel.7964....1D}, (17)~\citet{2015ATel.7965....1D}, (18)~\citet{2015ATel.7984....1H}, (19)~\citet{2016ATel.9848....1I}, (20)~\citet{12a2016}.}
\end{table}

These rapid-fire eruptions of \novak\ are powered by the most massive accreting WD yet discovered.  Studies of the 2013 eruption yielded $M_\mathrm{WD}>1.3\,M_\odot$ \citep[hereafter \ponek]{2014ApJ...786...61T}, with a more recent determination of $M_\mathrm{WD}=1.38\,M_\odot$ \citep{2015ApJ...808...52K}.  We note that the WD mass has not been measured directly, only estimated based on modeling of the system.  That same modeling required very large accretion rates, $\dot{M}_\mathrm{acc}>1.7\times10^{-7}\,M_\odot\,\mathrm{yr}^{-1}$ and $\dot{M}_\mathrm{acc}=1.6\times10^{-7}\,M_\odot\,\mathrm{yr}^{-1}$, respectively.  Under an assumption of spherical ejecta, \citet[hereafter \xtwok]{2015A&A...580A..46H} concluded that the quantity of ejected hydrogen was $M_\mathrm{e,H}=\left(2.6\pm0.4\right)\times10^{-8}\,M_\odot$, broadly consistent with the total ejected mass prediction of $M_\mathrm{e}=6\times10^{-8}\,M_\odot$ from \citet{2015ApJ...808...52K}.  This indicates a mass accretion efficiency of $\sim63\%$, not only is the WD massive, but it is growing.

\citet[hereafter \oonek]{2014A&A...563L...9D} and \ponek\ both illustrated the rapid optical development of the 2013 eruption; \citet[hereafter \xonek]{2014A&A...563L...8H} and \ponek\ noted the rapid X-ray development.  \othreek\ combined all data from the near-identical 2013, 2014, and 2015 eruptions to determine that the optical decay time ($t_2$; the time to decay two magnitudes from the peak luminosity) is only $1.65\pm0.04$\,days, and $t_3=2.47\pm0.06$\,days.  The accompanying super-soft X-ray source (SSS) `turned on' only $5.6\pm0.7$\,days after the 2015 eruption, and turned off after $18.6\pm0.7$\,days (\othreek); only the Galactic RN V745\,Scorpii displays more rapid X-ray evolution \citep{2015MNRAS.454.3108P}.

\othreek\ also presented a detailed analysis of the combined spectra of the 2012--2015 eruptions.  The earliest post-eruption spectra show fleeting evidence of very high velocity ($v_\mathrm{ej}\simeq13,000$\,km\,s$^{-1}$) outflows.  \othreek\ proposed that these could be due to a high level of ejecta collimation, in the polar direction, almost along the line of sight.  \citet[hereafter \hstk]{0004-637X-847-1-35} reported similar high velocity material surround the far UV N\,{\sc v} (1240\,\AA) emission line 3 days after the 2015 eruption, again this was linked to possible ejecta collimation or jets from the eruption. 

Hints of ejecta deceleration were first reported by \ponek.  The \othreek\ analysis of the combined 2012--2015 spectroscopy found clear evidence of significant ejecta deceleration, consistent with the adiabatic expansion of a forward shock \citep[cf.][]{1985MNRAS.217..205B}.  \otwok\ and \othreek\ both proposed that this deceleration could be caused by the ejecta interacting with pre-existing circumbinary material.  Given that the circumbinary regime should be cleared by each annual eruption, this environment must be regularly resupplied.  Therefore, \othreek\ proposed that the \novak\ donor should be a giant with a significant stellar wind, and not Roche lobe overflow. 

Utilising the {\it Swift} observatory, \citet{2016ApJ...830...40K} undertook the first targeted survey to detect the long-predicted X-ray flash precursor to a nova eruption \citep[see, e.g.,][]{1990LNP...369..306S,2002AIPC..637..345K}.  The campaign was unsuccessful, possibly because of the earlier than predicted 2015 eruption, or because the flash was absorbed by pre-existing material surrounding the system.  At the time, there was no strong evidence constraining the mass donor in the system, therefore \citet{2016ApJ...830...40K} favoured the former explanation. 

Containing a growing WD that is already close to the Chandrasekhar limit, \novak\ is therefore the leading pre-explosion supernova type Ia candidate system.  {\it Hubble Space Telescope} (HST) spectroscopy of the 2015 eruption conducted by \hstk\ found no evidence of neon within the ejecta.  However, as discussed by those authors, that single result cannot yet completely rule out the presence of an ONe WD in the system.  Either way, \hstk\ argued that the lack of an observational signature of Ne may in itself indicate that the \novak\ WD is growing in mass.  That is, either a CO WD has grown to the Chandrasekhar limit, or a large enough He layer has been accumulated to shield an underlying ONe WD from the nova eruptions. 

In this paper, we present the results of a HST program to study the late-decline of the predicted 2015 eruption of \novak, and an updated analysis of archival {\it HST} and Keck observations of the system.  In Section~\ref{sec:observations} we describe our observations, in Section~\ref{sec:phot} we present the photometric data.    In Section~\ref{sec:mod} we explore models of the accretion disk in \novak.  Finally, in Sections~\ref{sec:disc} and \ref{sec:conc} we discuss our findings and present the subsequent conclusions.

While this manuscript was being prepared, the 2016 eruption of \novak\ was detected by \citet{2016ATel.9848....1I}.  The  observations of the 2016 eruption will be presented in \citet{12a2016}.

\section{Observations}\label{sec:observations}

\subsection{{\textit Hubble Space Telescope} Observations}

Twenty orbits of {\it HST} Cycle\,23 time were awarded to collect early-time UV spectroscopic observations (8 orbits) and late-time imaging of the 2015 eruption of \novak\ (proposal ID: 14125).  The results of the spectroscopy were presented in \hstk. The 2015 eruption was discovered on 2015 Aug.\ 28.425\,UT by an automated monitoring program on the Las Cumbres Observatory 2\,m telescope\footnote{Formerly known as the Faulkes Telescope North} on Hawai'i (\citealp{2015ATel.7964....1D}, see \othreek\ for full details).  The {\it HST} photometric observations were conducted between 2015 Sep.\ 10 and Sep.\ 30, a log of these observations is provided in Table~\ref{obs_log}.

\begin{table*}
\caption{Log of observations of the eruptions of \novak\ referred to in this Paper.\label{obs_log}}
\begin{center}
\begin{tabular}{lllllllll}
\hline
\hline
Eruption & Facility & Instrument & {\it HST}  & Date & Start & End & Orbits & Exposure\\
     &   &    & Visit              & (midpoint) & \multicolumn{2}{c}{$t-t_0$ (days)} & & time (ks)\\
\hline
2014 & Keck I & LRIS & \nodata & 2014 Oct 21.50 & 18.80 & 18.82 & \nodata & \phn1.2 \\
\hline
2015 & {\it HST} & WFC3/UVIS & 4 & 2015 Sep 10.64 & 13.27 & 13.44 & 3 & \phn6.8\\
2015 & {\it HST} & WFC3/UVIS & 5 & 2015 Sep 17.66 & 20.29 & 20.47 & 3 & \phn6.8\\
2015 & {\it HST} & WFC3/UVIS & 6 & 2015 Sep 23.62 & 26.26 & 26.43 & 3 & \phn6.8\\
2015 & {\it HST} & WFC3/UVIS & 7 & 2015 Sep 30.58 & 33.22 & 33.39 & 3 & \phn6.8\\
\hline
\end{tabular}
\end{center}
\end{table*}

We employed 12 {\it HST} orbits, split into four visits, to collect photometry of \novak\ using the Wide Field Camera 3 (WFC3) in the UVIS mode.  Each visit used identical observing strategies and were approximately one week apart, starting at $\Delta t\simeq 14$\,days (post-eruption).  Observations were obtained using the WFC3/UVIS F225W, F275W, F336W, F475W, and F814W filters.

For each filter a two-point dither was applied to enable removal of detector defects. To reduce readout overheads, WFC3/UVIS was operated in a $2\mathrm{k}\times2\mathrm{k}$ windowed mode utilizing the UVIS2-2K2C-SUB aperture. This part of the chip was selected for its superior performance against charge transfer efficiency (CTE) loss, to further mitigate such effects we included a `post flash' signal of 9--12 electrons.

The WFC3/UVIS data were reduced using the STScI {\tt calwf3} pipeline \citep[v3.1.6; see][]{2012wfci.book.....D}, with CTE correction manually applied via the {\tt wfc3uv\_ctereverse\_parallel} code \citep[v2015.07.22\footnote{\url{http://www.stsci.edu/hst/wfc3/tools/cte_tools}}; see also][]{CTE}. Photometry of the WFC3/UVIS data was then performed on individual exposures using DOLPHOT \citep[v2.0\footnote{\url{http://americano.dolphinsim.com/dolphot}};][following the standard procedure and parameters for WFC3/UVIS given in the manual]{2000PASP..112.1383D}. For comparative purposes, photometry was also carried out using the combined exposures per epoch for each filter. All data were aligned and final combined images created using the Drizzlepac (v2.0.2) {\tt astrodrizzle} package. Photometry was obtained via the PyRAF {\tt phot} package (v2.2). The results from the DOLPHOT and {\tt phot} methods are consistent.  For comparison with previous work, we adopt the DOLPHOT photometry, which is presented in Table~\ref{phot_table}. 

\begin{table*}
\caption{{\it Hubble Space Telescope} WFC3/UVIS NUV and visible photometry of \novak\ following the late decline of the 2015 eruption.\label{phot_table}}
\begin{center}
\begin{tabular}{llllll}
\hline
\hline
Date & $\Delta t$ & Exposure & Filter & S/N & Photometry \\
(UT) & (days) & (secs) & & & (Vega mag)\\
\hline 
2015-09-10.669 & $13.389\pm0.027$  & $2\times870$ & F225W & \phn55.0 & $20.83\pm0.02$ \\
2015-09-17.693 & $20.413\pm0.027$  & $2\times870$ & F225W & \phn36.5 & $21.83\pm0.03$ \\
2015-09-23.656 & $26.376\pm0.027$  & $2\times870$ & F225W & \phn25.8 & $22.31\pm0.04$ \\
2015-09-30.333 & $33.333\pm0.027$  & $2\times870$ & F225W & \phn21.2 & $22.83\pm0.05$ \\
\hline
2015-09-10.564 & $13.284\pm0.007$  & $2\times519$ & F275W & \phn38.5 & $20.80\pm0.03$ \\
2015-09-17.587 & $20.307\pm0.007$  & $2\times519$ & F275W & \phn25.1 & $22.18\pm0.04$ \\
2015-09-23.550 & $26.270\pm0.007$  & $2\times519$ & F275W & \phn19.9 & $22.52\pm0.06$ \\
2015-09-30.228 & $33.228\pm0.007$  & $2\times519$ & F275W & \phn18.4 & $22.79\pm0.06$ \\
\hline
2015-09-10.580 & $13.300\pm0.007$  & $2\times519$ & F336W & \phn76.0 & $21.08\pm0.01$ \\
2015-09-17.603 & $20.324\pm0.007$  & $2\times519$ & F336W & \phn47.2 & $22.08\pm0.02$ \\
2015-09-23.566 & $26.286\pm0.007$  & $2\times519$ & F336W & \phn35.1 & $22.61\pm0.03$ \\
2015-09-30.245 & $33.245\pm0.007$  & $2\times519$ & F336W & \phn33.2 & $22.75\pm0.03$ \\
\hline
2015-09-10.708 & $13.428\pm0.010$  & $2\times745$ & F475W & 130.7 & $22.48\pm0.01$ \\
2015-09-17.733 & $20.453\pm0.010$  & $2\times745$ & F475W & \phn83.3 & $23.56\pm0.01$ \\
2015-09-23.696 & $26.416\pm0.010$  & $2\times745$ & F475W & \phn65.4 & $23.97\pm0.02$ \\
2015-09-30.372 & $33.372\pm0.010$  & $2\times745$ & F475W & \phn60.1 & $24.18\pm0.02$ \\
\hline
2015-09-10.629 & $13.349\pm0.010$  & $2\times765$ & F814W & \phn85.7 & $22.37\pm0.01$ \\
2015-09-17.654 & $20.374\pm0.010$  & $2\times765$ & F814W & \phn47.9 & $23.39\pm0.02$ \\
2015-09-23.617 & $26.337\pm0.010$  & $2\times765$ & F814W & \phn34.7 & $23.84\pm0.03$ \\
2015-09-30.293 & $33.293\pm0.010$  & $2\times765$ & F814W & \phn34.9 & $23.91\pm0.03$ \\
\hline
\end{tabular}
\end{center}
\end{table*}

\subsection{Keck spectroscopy of the 2014 eruption}

\othreek\ observed that photometrically, the 2013, 2014, and 2015 eruptions were essentially identical; the same is true of the spectra from the 2012--2015 eruptions.  Therefore, to support the late-time HST photometry of the 2015 eruption, we also utilise a Keck spectrum of the 2014 eruption taken 18.81\,days after that eruption.

This 2014 Keck spectrum has not been published until now.  It was collected using the Low Resolution Imaging Spectrometer \citep[LRIS;][]{1995PASP..107..375O,1998SPIE.3355...81M,2010SPIE.7735E..0RR}, which is mounted at the Cassegrain focus of the Keck I telescope on Mauna Kea, Hawai'i.  The spectrum was obtained through the standard low-resolution configuration using the 400/3000 grism (blue camera) and 400/8500 grating (red camera), providing continuous coverage from the atmospheric cutoff to approximately 10300\,\AA.  However, as the nova had faded significantly, crowding and confusion with nearby stars in M31 had started to be problematic, therefore, the object is only clearly detected in the blue camera; only data with $\lambda < 5600$\,\AA\ are analyzed here. 
 
 \subsection{Archival Quiescent Data}
 
The Panchromatic Hubble Andromeda Treasury \citep[PHAT;][]{2012ApJS..200...18D} was a broadband, multicolour, NUV--NIR {\it HST} survey of the bulge and north-eastern disk of M31. As part of the PHAT survey, \novak\ was observed between eruptions a number of times with {\it HST}.  Initial results from analysis of these data were published in \oonek\ and \ponek.  Both those works analyzed the optical and NUV {\it HST} data finding evidence for a very blue source coincident with \novak, indicating the presence of a luminous accretion disk.  Although the available {\it HST} NIR data were also analyzed, \oonek\ and \ponek\ only presented upper limits on the quiescent photometry of \novak, which was severely blended with nearby sources in the NIR.  These upper limits did not place firm constraints on the nature of the donor, only excluding the most luminous red giants (such as that found in the T\,Coronae Borealis system).  Notably, the initial analysis of the quiescent SED indicated an accretion disk roughly similar in flux distribution, albeit brighter, to that in the RS\,Oph system; a donor of similar luminosity to the red giant in RS\,Oph was not ruled out by \oonek.

\citet{2014ApJS..215....9W} released the NUV to NIR photometric catalog from the PHAT survey, which included the quiescent photometry of \novak.  These photometry are provided in Table~\ref{PHAT_phot} and are consistent with the independent analysis by \oonek\ and \ponek.  However, the analysis undertaken by \citet{2014ApJS..215....9W} was able to successfully de-blend the sources around \novak\ in the NIR yielding F110W and F160W photometry of the quiescent system.  This superior NIR deblending was achieved by simultaneous fitting of the higher spatial resolution F475W data with the NIR data.  These F475W data have spatial resolution better by more than a factor of two and allowed for much more robust deblending of crowded sources.

\begin{table*}
\caption{PHAT multicolor NUV, optical, and NIR photometry of \novak, in part from \citet{2014ApJS..215....9W}.\label{PHAT_phot}}
\begin{center}
\begin{tabular}{lllllllll}
\hline
\hline
Date & \multicolumn{2}{c}{Observed eruptions} & \multicolumn{2}{c}{Predicted eruptions} & {\it HST} & Filter & Exposure & Photometry \\
(UT) & $\Delta t_{\mathrm{after}}$ & $\Delta t_{\mathrm{before}}$ & $\Delta t_{\mathrm{after}}$ & $\Delta t_{\mathrm{before}}$ & Instrument && Time \\
 & (days) & (days) & (days) & (days) &&& (s)\\
\hline
  2011 Jan.\ 25.21 & \phn67 & 270 & \phn67 & $115\pm26$ & WFC3/UVIS & F275W & 350 & $23.13\pm0.12$\tablenotemark{\dag}\\
  2011 Jan.\ 25.23 & \phn67 & 270 & \phn67 & $115\pm26$ & WFC3/UVIS & F275W & 660 & $22.98\pm0.07$\tablenotemark{\dag}\\
   2011 Aug.\ 31.51 & 285 & \phn52 & $103\pm26$ & \phn52 & WFC3/UVIS & F275W & 350 & $22.73\pm0.09$\\
 2011 Aug.\ 31.53 & 285 & \phn52 & $103\pm26$ & \phn52 & WFC3/UVIS & F275W & 575 & $22.53\pm0.06$\\
\hline
  2011 Jan.\ 25.20 & \phn67 & 270 & \phn67 & $115\pm26$ & WFC3/UVIS & F336W & 550 & $23.07\pm0.05$\tablenotemark{\dag}\\
  2011 Jan.\ 25.22 & \phn67 & 270 & \phn67 & $115\pm26$ & WFC3/UVIS & F336W & 800 & $23.01\pm0.04$\tablenotemark{\dag}\\
 2011 Aug.\ 31.51 & 285 & \phn52 & $103\pm26$ & \phn52 & WFC3/UVIS & F336W & 550 & $22.59\pm0.04$\\
   2011 Aug.\ 31.52 & 285 & \phn52 & $103\pm26$ & \phn52 & WFC3/UVIS & F336W & 700 & $22.59\pm0.04$\\
\hline
  2010 Aug.\ 07.53 & 248 & 104 & \phn$68\pm26$ & 104 & ACS/WFC & F475W & 600 & $24.08\pm0.02$\tablenotemark{\dag}\\
  2010 Aug.\ 07.53 & 248 & 104 & \phn$68\pm26$ & 104 & ACS/WFC & F475W & 370 & $24.06\pm0.03$\tablenotemark{\dag}\\
  2010 Aug.\ 07.54 & 248 & 104 & \phn$68\pm26$ & 104 & ACS/WFC & F475W & 370 & $24.01\pm0.03$\tablenotemark{\dag}\\
  2010 Aug.\ 07.54 & 248 & 104 &\phn $68\pm26$ & 104 & ACS/WFC & F475W & 370 & $24.08\pm0.03$\tablenotemark{\dag}\\
   2012 Jan.\ 10.12 & \phn80 & 282 & \phn80 & $118\pm26$ &  ACS/WFC & F475W & 700 & $24.46\pm0.03$\\
 2012 Jan.\ 10.13 & \phn80 & 282 & \phn80 & $118\pm26$ &  ACS/WFC & F475W & 360 & $24.48\pm0.04$\\
 2012 Jan.\ 10.13 & \phn80 & 282 & \phn80 & $118\pm26$ &  ACS/WFC & F475W & 360 & $24.43\pm0.04$\\
 2012 Jan.\ 10.14 & \phn80 & 282 & \phn80 & $118\pm26$ &  ACS/WFC & F475W & 470 & $24.51\pm0.03$\\
\hline
  2010 Aug.\ 07.45 & 248 & 104 & \phn$68\pm26$ & 104 & ACS/WFC & F814W & 350 & $23.87\pm0.05$\tablenotemark{\dag}\\
  2010 Aug.\ 07.46 & 248 & 104 & \phn$68\pm26$ & 104 & ACS/WFC & F814W & 700 & $23.80\pm0.03$\tablenotemark{\dag}\\
  2010 Aug.\ 07.47 & 248 & 104 & \phn$68\pm26$ & 104 & ACS/WFC & F814W & 455 & $23.83\pm0.04$\tablenotemark{\dag}\\
 2012 Jan.\ 10.02 & \phn80 & 282 & \phn80 & $118\pm26$ & ACS/WFC & F814W & 350 & $23.98\pm0.05$\\
 2012 Jan.\ 10.05 & \phn80 & 282 & \phn80 & $118\pm26$ & ACS/WFC & F814W & 800 & $23.97\pm0.04$\\
 2012 Jan.\ 10.06 & \phn80 & 282 & \phn80 & $118\pm26$ & ACS/WFC & F814W & 550 & $23.99\pm0.04$\\
\hline
  2011 Jan.\ 25.27 & \phn67 & 270 & \phn67 & $115\pm26$ & WFC3/IR & F110W & 800 & $24.19\pm0.05$\tablenotemark{\dag}\\
   2011 Aug.\ 31.58 &  285 &  \phn52 &  $103\pm26$ &  \phn52 & WFC3/IR & F110W & 700 & $23.71\pm0.03$\\
\hline
  2011 Jan.\ 25.26 & \phn67 & 270 & \phn67 & $115\pm26$ & WFC3/IR & F160W & 400 & $24.1\phn\pm0.2$\tablenotemark{\dag}\\
  2011 Jan.\ 25.28 & \phn67 & 270 & \phn67 & $115\pm26$ & WFC3/IR & F160W & 400 & $24.0\phn\pm0.2$\tablenotemark{\dag}\\
  2011 Jan.\ 25.29 & \phn67 & 270 & \phn67 & $115\pm26$ & WFC3/IR & F160W & 400 & $23.9\phn\pm0.2$\tablenotemark{\dag}\\
  2011 Jan.\ 25.29 & \phn67 & 270 & \phn67 & $115\pm26$ & WFC3/IR & F160W & 500 & $24.2\phn\pm0.2$\tablenotemark{\dag}\\
 2011 Aug.\ 31.57 &  285 & \phn52  &  $103\pm26$ &  \phn52 & WFC3/IR & F160W & 400 & $23.5\phn\pm0.1$\\
   2011 Aug.\ 31.59 &  285 & \phn52  &  $103\pm26$ &  \phn52 & WFC3/IR & F160W & 400 & $23.5\phn\pm0.1$\\
 2011 Aug.\ 31.59 &  285 & \phn52  &  $103\pm26$ &  \phn52 & WFC3/IR & F160W & 400 & $23.4\phn\pm0.1$\\
 2011 Aug.\ 31.60 &  285 & \phn52  &  $103\pm26$ &  \phn52 & WFC3/IR & F160W & 400 & $23.3\phn\pm0.1$\\
\hline
\end{tabular}
\end{center}
\tablenotetext{\dag}{Data derived directly from \cite{2014ApJS..215....9W}; the remainder have been provided directly by the PHAT collaboration.}
\end{table*}

The observations reported in \citet{2014ApJS..215....9W} are from the first set of PHAT visits and are computed over two separate {\it HST} visits.  Data from another pair of visits are also available and the PHAT collaboration have generously supplied their photometry of \novak\ from each of the four {\it HST} visits, and these data are also shown in Table~\ref{PHAT_phot}.  The quiescent photometry reported by \oonek\ and \ponek\ are combined from observations at different phases in the full eruption cycle of \novak, and from different eruption cycles (as noted by both those papers).   

\section{Light curve analysis}\label{sec:phot}

\begin{figure*}
\includegraphics[width=\textwidth]{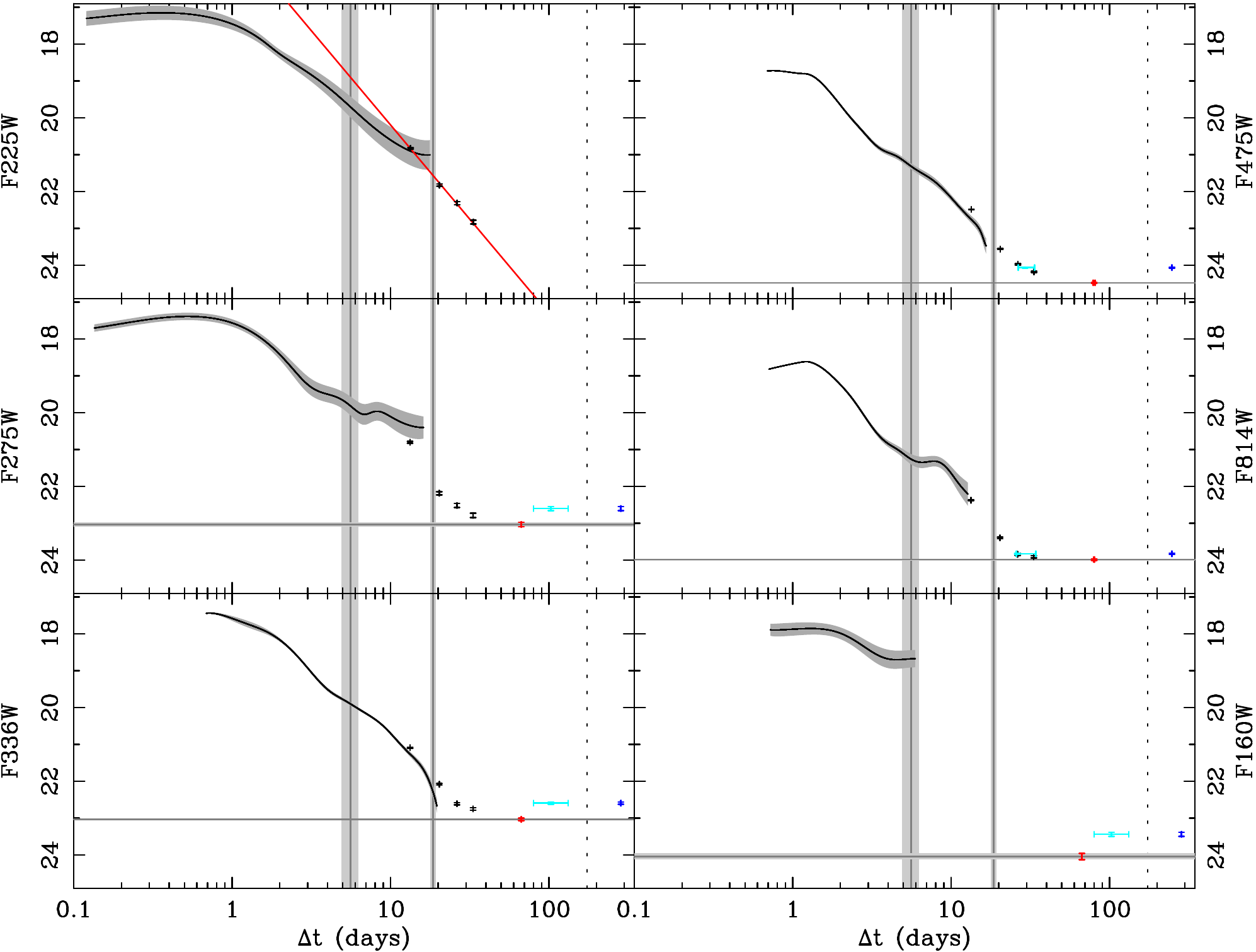}
\caption{Near-ultraviolet through optical {\it HST} WFC3/UVIS photometry of the 2015 eruption of \novak\ (black data points).  The time axes runs from 0.1\,days post-eruption up to 347\,days -- the mean observed recurrence period.  The vertical dashed line indicates the proposed 147\,day recurrence period (\halfk).  The horizontal lines, where shown, indicate the faintest detection of the two PHAT epochs -- assumed to be the quiescence level.  The vertical gray lines indicate the turn on and turn off times of the SSS from the 2015 eruption (the shaded areas their associated uncertainties).  The solid black lines show combined and smoothed (B\'{e}zier curve) photometry from the 2013, 2014, and 2015 eruptions of \novak, with the surrounding shaded area indicating the $1\sigma$ uncertainty; these smoothed lines are provided for illustrative and contextual purposes only, the {\it HST} and ground-based/{\it Swift} data are taken through similar, but different, filters.  The solid red line in the F225W shows the best fit power-law of index $-2.04\pm0.16$.  The red data points indicate the faintest archival PHAT photometry following a detected eruption, the dark-blue points indicate additional PHAT photometry (assuming a year-long cycle), the light-blue points are the same PHAT photometry points extrapolated to a predicted missed eruption (based on a six month recurrence period).\label{log_lc}}
\end{figure*}

The five-band {\it HST} photometry was presented in Table~\ref{phot_table}, and the subsequent light curves are presented in Figure~\ref{log_lc}.  In the sub-figures, the four epochs of {\it HST} WFC3/UVIS observations (black data points) are compared with the template \novak\ eruption light curves from other telescopes.  These eruption templates are constructed from B\'{e}zier smoothed light curves of the almost identical 2013--2015 eruptions (\oonek, \otwok, \othreek).  The uncertainties on the smoothed light curves are computed based on the method employed by \citet{2016MNRAS.460.3529A}.  Here, we compare to the closest filter in wavelength to the {\it HST} filters.  The F275W filter is compared to the {\it Swift} UVW1 data (central wavelength 2600\,\AA), F336W to Sloan $u'$-band (the Sloan ground-based data are converted from the AB system to the Vega system in this plot), F475W to $B$, and F814W to $i'$, the {\it Swift} UVM2 filter (2250\,\AA) is used for comparison to the F225W data, but these data are not particularly extensive.  The solid vertical lines in each plot indicate the epochs of the SSS turn-on and turn-off.  The horizontal lines indicate the minimum photometry from the two visits of the PHAT survey (see Table~\ref{PHAT_phot}).

The {\it HST} WFC3/UVIS F225W data are well fit by a power law of the form $f\propto t^\alpha$, where $\alpha=-2.04\pm0.16$ ($\chi^{2}_{/\mathrm{dof}}=2.2$). We note that this decline is therefore consistent with the `middle' relation predicted by the universal decline law of \citet[$\alpha=-1.75$]{2006ApJS..167...59H,2007ApJ...662..552H}.  This may be connected to the lack of strong emission lines seen in the equivalent region of the NUV spectrum (see \hstk). We also note that \othreek\ found that a power-law fit to the $u'$-band decline of the 2015 eruption, between days 8 and 20, was consistent with the predicted `middle' decline law.  The {\it HST} data from the other filters all show significant deviation from a {\it single} power law, when taken in isolation, and when compared to the eruption template data -- all these filters contain strong flux contributions from lines.

\subsection{Quiescent Data}\label{aqd}

Comparison between the {\it HST} imaging of the 2015 eruption and the archival data confirm that the object proposed as the quiescent system by \oonek\ and \ponek\ is associated with the eruptions of \novak. In Table~\ref{PHAT_phot} we have also indicated the epoch of the archival PHAT {\it HST} visits in respect to the \novak\ eruption cycle. The closest PHAT observations to a known eruption are those from 2011 Jan., which took place 67\,days after the 2010 Nov.\ eruption; significantly later, post-eruption, than the late-time decline data collected for this paper.  We also note that the 2011 Aug.\ observations took place 52 days before the 2011 Oct.\ eruption.

\halfk\ presented evidence that \novak\ may erupt every $\sim6$\,months, rather than annually.  If this is the case, we must also assess whether the interpretation of the archival {\it HST} data may be affected by {\it unobserved} eruptions.  The typical eruption date uncertainty is 26 days (\halfk).  If we utilize the dates of observed eruptions but assume a $\sim6$\,month cycle (see \halfk), we can investigate how close to an {\it unobserved} eruption each PHAT visit potentially occured (recorded in Table~\ref{PHAT_phot}).  The only observations of note here are those from 2010 Aug.\ which may lie $68\pm26$\,days after an unobserved {\it early} 2010 eruption.  The F475W and F814W data from that time are significantly brighter than those from 2012 Jan.\ (80 days after the 2011 Oct.\ eruption), which suggests that these data {\it may} be coincident with the late decline of an {\it early} (but missed) 2010 eruption.

By assuming that all \novak\ eruptions are essentially identical we can roughly fit the 2010 Aug.\ {\it HST} observations to the 2015 eruption late decline observations. Therefore, we would predict that a missed eruption of \novak\ could have occurred on 2010 Jul.\ 09$^{+4}_{-3}$ (see the light-blue data points in Figure~\ref{log_lc}).  However, data from PTF rule out an additional eruption between 2010 Jun.\ 30 and the date of the {\it observed} 2010 Nov.\ eruption \citep[M.\ M.\ Kasliwal priv.\ comm.]{2012ApJ...752..133C}.  As such, we conclude that all  PHAT data of \novak\ were taken at least 67\,days after an eruption, and that they represent observations of the inter-eruption, or quiescent, period.   We stress that this does not rule out the possibility of an {\it early} 2010 eruption occurring before this window.

\subsection{A `folded' eruption cycle}

The {\it HST} data covering the quiescent system are admittedly sparse and spread across multiple eruption cycles.  However, under the assumption of essentially identical eruptions (\citealt{2010ApJS..187..275S}, \othreek), \novak\ appears to take $\sim70$\,days to return to quiescence, i.e., to reach a minimum flux following an eruption.  From this point, the luminosity of the system appears to increase in the lead-up to the next eruption, consistent with the findings of \citet{HenzeQ}.

The RN RS\,Oph is perhaps the best studied Galactic nova both during eruption and at quiescence \citep[see][and references therein]{2008ASPC..401.....E}.  Following the 2006 eruption of RS\,Oph the system was observed to decline to an optical minimum before the flux began to systematically increase.  The increase in flux was more prominent in bluer bands \citep{2008ASPC..401..203D}, and coincided with the resumption of optical flickering \citep{2007MNRAS.379.1557W}.  These observations were proposed to indicate the re-establishment of accretion post-eruption -- following the destruction or severe disruption of that disk.

By mapping the quiescent PHAT data onto the template light curves we can combine these multi-color data into two distinct quiescent epochs, based on their approximate phase in the eruption cycle.  The first (the red points in Figure~\ref{log_lc}) $\sim75$\,days post eruption represents the approximate minimum luminosity state, the second (dark-blue points) $\sim270$\,days post eruption shows a state of increased flux.  We again note that the lighter-blue data points in Figure~\ref{log_lc} indicate one possible realization of a six month recurrence period; a realization that is ruled out by PTF data (see Section~\ref{aqd}).

\subsection{Spectral Energy Distribution}\label{sed_sec}

Optical and NUV photometric observations of the 2014 eruption of \novak\ indicated that, due to the low ejected mass, the unusually low maximum radius of the expanding pseudo-photosphere resulted in emission peaking in the UV (\otwok).  For all other well observed novae, this peak occurs at visible wavelengths.

\begin{figure*}
\includegraphics[width=\columnwidth]{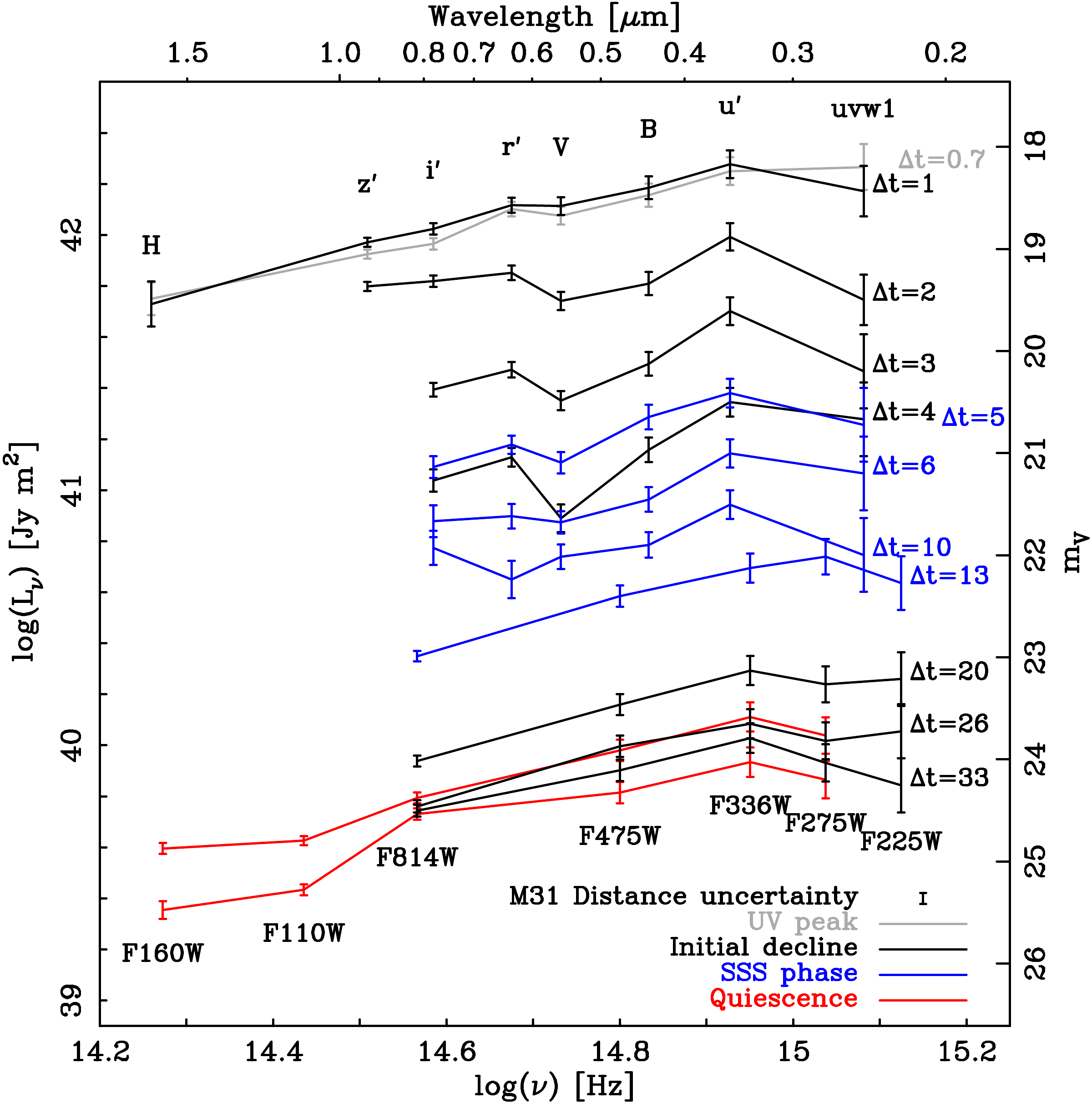}\hfill
\includegraphics[width=\columnwidth]{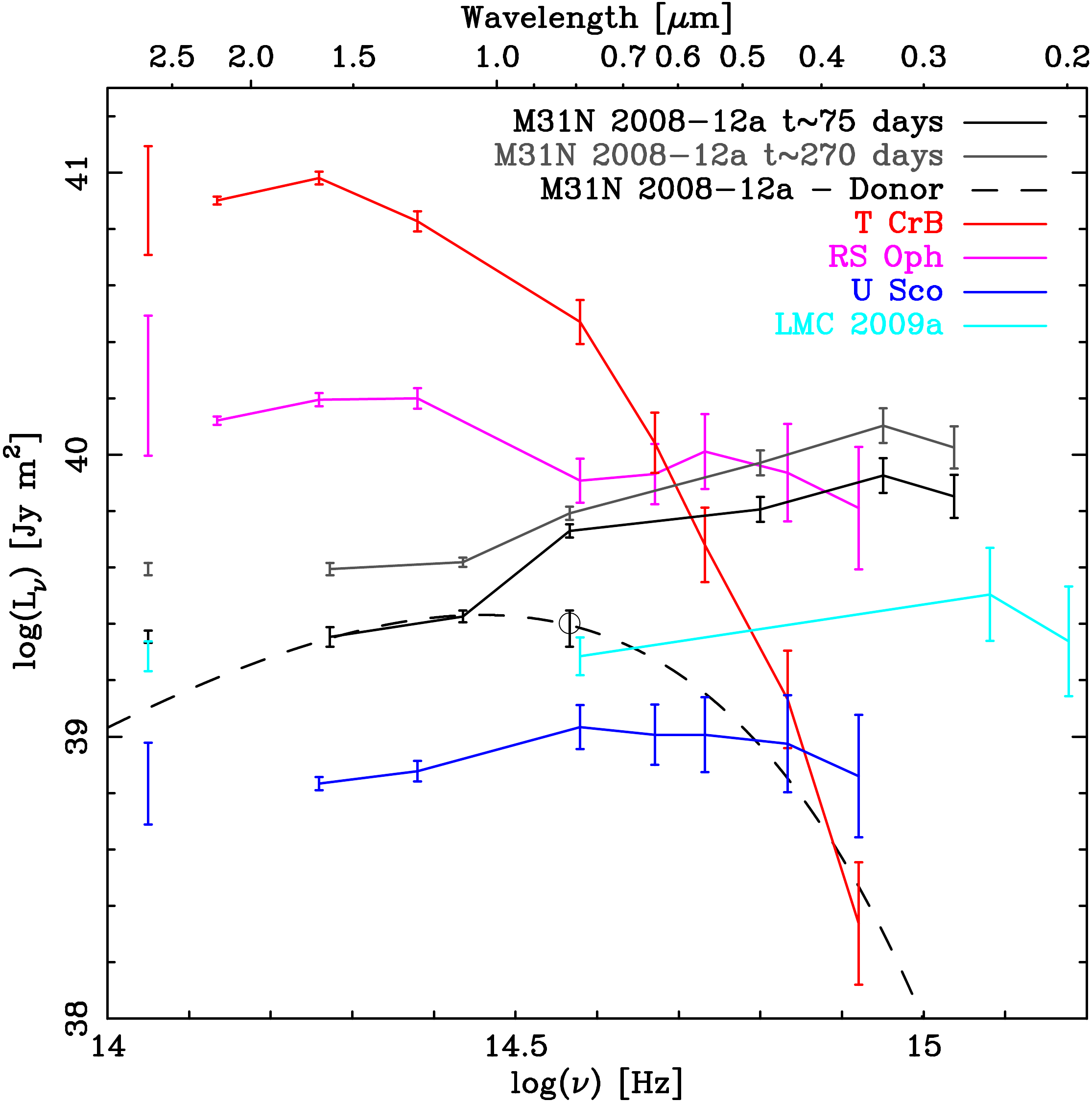}
\caption{Distance and extinction ($E_{B-V}=0.1$; \hstk) corrected SEDs showing {\bf Left:} the evolving SED of the 2015 eruption (blue points indicate epochs when the SSS emission was visible, red points indicate the archival photometry), and {\bf Right:} the quiescent \novak\ compared to the quiescent RNe RS\,Oph, T\,CrB, U\,Sco, and LMC\,2009a. The black data shows \novak\ at minimum ($\sim75$\,days post-eruption), and the gray data at an elevated state ($t\simeq270$\,d).  Throughout, units chosen to allow comparison with similar plots in \citet[see their Fig.\ 71]{2010ApJS..187..275S} and \oonek\ (see their Fig.\,4).  The central wavelength locations of
the Johnson-Cousins, Sloan, {\it HST}, and {\it Swift} filters are shown to assist the reader, see the Keys for line identifications.  For each system, the photometric uncertainties are relatively small, indicated error bars are dominated by extinction uncertainties, the isolated error bar to the left of each SED indicates the systematic distance uncertainty.  The single circled black data point indicates the excess donor flux once the accretion disk model (see Section~\ref{donor_excess}) has been subtracted.  The dashed black line indicates a power law-fit to the \novak\ donor SED (see Section~\ref{sec:donor}).\label{fig:sed_q}\label{fig:sed_evo}}
\end{figure*}

\othreek\ presented a more comprehensive series of SEDs following the NIR ($H$-band) through NUV ({\it Swift} uvw1) decline of the 2015 eruption of \novak\ spanning $t\sim1-10$\,days post-eruption.  In Figure~\ref{fig:sed_evo} we reproduce the SED evolution plot from \othreek\ and include the {\it HST} WFC3/UVIS photometry from the 2015 eruption of \novak\ ($t\sim13$, 20, 26, and 33\,days post-eruption).  We also include the updated quiescent photometry from archival {\it HST} observations.  In Section~\ref{sec:mod} we will use these new data in conjunction with model accretion disks to constrain the mass accretion rates.  The nature of the quiescent system is explored in Section~\ref{sec:donor}.

\section{Modeling the accretion disk}\label{sec:mod}

In this Section, and subsequently in Section~\ref{sec:disc}, we discuss in detail our models and interpretation of the accretion disk in \novak.  Here, for clarity, we formally define some of the accretion rate terminology that we employ.

The models, discussed below, generate the disk mass accretion rate ($\dot{M}$), whereas the existing \novak\ eruption models of \citet{2014ApJ...793..136K,2015ApJ...808...52K,2016ApJ...830...40K,2017ApJ...838..153K,2017ApJ...844..143K} are concerned with the WD mass accretion rate ($\dot{M}_\mathrm{acc}$); the amount of material that falls onto the WD surface itself.

In this work we will also consider mass loss from the disk via a disk wind ($\dot{M}_\mathrm{wind}$), and mass loss from any outflows from the WD or the disk--WD boundary layer ($\dot{M}_\mathrm{bl}$); such that:

\[
\dot{M}_\mathrm{acc}=\dot{M}-\dot{M}_\mathrm{wind}-\dot{M}_\mathrm{bl},
\]

\noindent for most novae $\dot{M}$ is low, therefore it is expected that $\dot{M}_\mathrm{wind}-\dot{M}_\mathrm{bl}$ are small, and as such $\dot{M}_\mathrm{acc}\simeq\dot{M}$.

\subsection{Disk Models}
              
The {\tt tlusty}, {\tt synspec}, {\tt rotin}, and {\tt disksyn} suite of codes \citep{1988CoPhC..52..103H,hub94,1995ApJ...439..875H} are employed to generate synthetic spectra of stellar atmospheres and disks.  These include the treatment of hydrogen quasi-molecular satellite lines (low temperature) and NLTE approximation (high temperature). {\tt synspec} generates continuum spectra with absorption lines.  In the present work we do not generate emission lines \cite[see, e.g.,][for a physical description of emission line profiles from disks in CVs]{2007AJ....134.1923P}.  For disk spectra we assume Solar abundances and for stellar spectra we vary the abundances as required.  
                            
The {\tt tlusty} code is first run to generate one-dimensional (vertical) stellar atmosphere structures for a given surface gravity, effective temperature and surface composition of the star.   H and He are treated explicitly, whereas C, N, and O are treated implicitly \citep{1995ApJ...439..875H}.  
                
The {\tt synspec} code takes the {\tt tlusty} stellar atmosphere model as an input, and generates a synthetic stellar spectrum over a given wavelength range from below 900\,\AA\ and into the optical.   The {\tt synspec} code then derives the detailed radiation and flux distribution of the continuum and lines, to  generate the output spectrum \citep{1995ApJ...439..875H}.  {\tt synspec} has its own chemical abundances input to generate lines for the chosen species.  For temperatures $>$35000\,K the approximate NLTE line treatment is turned on in {\tt synspec}. 
   
Rotational and instrumental broadening, as well as limb darkening \citep[see][]{1998ApJ...509..350W}, are then reproduced using the {\tt rotin} routine.  In this manner, we generated WD synthetic spectra covering a wide range of temperatures and gravities, all with Solar composition. 
        
The disk spectra are generated by dividing the disk into annuli, with radius $r_i$ and effective surface temperature $T\left(r_i\right)$ obtained from the standard disk model for a given WD mass $M_\mathrm{WD}$ and mass loss rate $	$.  
      
Utilizing input parameters of the disk mass accretion rate ($\dot{M}$), $M_\mathrm{WD}$, the radius of the WD $R_\mathrm{WD}$, the inner radius of the disk $R_0$, and the outer radius of the disk $R_\mathrm{disk}$, {\tt tlusty} generates a one-dimensional vertical structure for each disk annulus \citep{1998ApJ...509..350W}.   

In the standard disk model, the radius $R_0$ is the boundary at which the `no shear' condition is imposed; $d\Omega / dR=0$ \citep{1977MNRAS.178..195P}.  Consequently, the assumed value of $R_0$ affects the entire solution (not just the boundary) and the temperature profile of the disk.  

For moderate disk mass accretion rates, $\dot{M} \sim 10^{-8}\,M_\odot$\,yr$^{-1}$, the boundary layer between the disk and the WD, that region where the angular velocity in the disk decreases from its Keplerian value $\Omega_K$ to match the more slowly rotating WD surface $\Omega_\star$, is very small ($\sim 0.01\,R_\mathrm{WD}$) and one can therefore assume $R_0 = R_\mathrm{WD}$ \citep{1977MNRAS.178..195P}.  

In our present modeling, $R_0$ is allowed to be larger than the radius of the WD, $R_0 > R_\mathrm{WD}$, to accommodate a larger boundary layer \citep[see][for a description of this modified disk model]{2017ApJ...846...52G}. As $\dot{M}$ increases, the boundary  layer becomes larger \citep{1995ApJ...442..337P}.  As $\dot{M}$ reaches the Eddington accretion limit, the size of the boundary layer rises to the order of the radius of the WD \citep[$R_0\sim R_\mathrm{WD}$;][]{1997ApJ...483..882G}.  

Given the large quiescent luminosity and high ejection velocities, \othreek\ and \hstk\ proposed that the system inclination must be low.  Although high inclination systems are not formally ruled out, we note that the large observed disk luminosity would require a significant increase in any derived $\dot{M}$ as the assumed inclination increases.

To model the \novak\ disk, we assume $M_\mathrm{WD}=1.37\,M_{\odot}$ and  $R_\mathrm{WD}=2000$\,km, yielding an Eddington limit $\dot{M}_\mathrm{Edd}=4 \times 10^{-6}\,M_\odot\,$\,yr$^{-1}$. We generate a grid of disk models for inclinations $i=10^{\circ}$, $20^{\circ}$, and $30^{\circ}$.  These models are computed for fixed values of $\dot{M}$ in logarithmic intervals of $0.5$.  The true value of $\dot{M}$ is computed by fitting the observed data by interpolating between the computed values of $\dot{M}$. 
For $R_0$, we choose: 

\[
R_0=
\begin{cases}
1.0 R_\mathrm{WD}, & \dot{M} \leq 10^{-7}\,M_\odot\,\mathrm{yr}^{-1}\\
1.1 R_\mathrm{WD}, & \dot{M} = 10^{-6.5}\ \mathrm{and}\ 10^{-6}\,M_\odot\,\mathrm{yr}^{-1}\\
1.5 R_\mathrm{WD}, & \dot{M} = 10^{-5.5}\,M_{\odot}\,\mathrm{yr}^{-1}\\
2.0 R_\mathrm{WD}, &  \dot{M} = 10^{-5}\,M_{\odot}\,\mathrm{yr}^{-1}.
\end{cases}
\]
 
For \novak, we use Kurucz stellar spectra of appropriate temperature and surface gravity to extend the outer disk to a radius where $3500\lesssim T < 10000$\,K. We also consider disks that are truncated in the outer region as discussed in the results section.  

{\tt synspec} uses the {\tt tlusty} results for each disk annulus to generate synthetic spectra.  These are integrated into a disk spectrum using {\tt disksyn}, which includes effects of Keplerian broadening, inclination, and limb darkening \citep{1998ApJ...509..350W}
   
\subsection{Results}

Here, we adopt an inclination of $20^{\circ}$, distance of 770\,kpc, and reddening $E_{B-V}=0.1$ (\hstk). In Section~\ref{model_uncert}, we take the effects of a different inclination ($10^{\circ}$ or $30^{\circ}$), an error of $\sim$20\,kpc on the distance, and a reddening error of 0.03 (\hstk) into consideration and assess how these affect the final results. Since the error bars on the data points are themselves rather small (at most 5\%), they too are considered at the end of this section.    A low inclination system is assumed due to the large UV flux at quiescence, and the large observed ejecta velocities (\othreek, \hstk).
 
The flux data points from the different epochs were obtained through filters covering given wavelength bands, and as such they represent an average continuum flux level in these regions of the spectrum, possibly also including some prominent lines. One data point (F475W; 4773.7\,\AA) includes H$\beta$ (which would be in absorption unless there is a disk wind, which we do not model here).  We therefore do not expect the data points, at any epoch, to line up nicely with the continuum of the optically thick standard disk model, {\it but rather we use our modeling simply to assess the order of magnitude of the mass accretion rate.  }

\subsubsection{Quiescence} 
  
We start by modeling the inter-eruption data at $t\simeq75$\,d, as here the flux is at a minimum and we expect the disk to dominate the optical--NUV emission, with negligible contribution from the waning eruption.  The modeling at this epoch is then applied to, and adjusted as necessary, the other five epochs.

\begin{figure}
\includegraphics[width=\columnwidth]{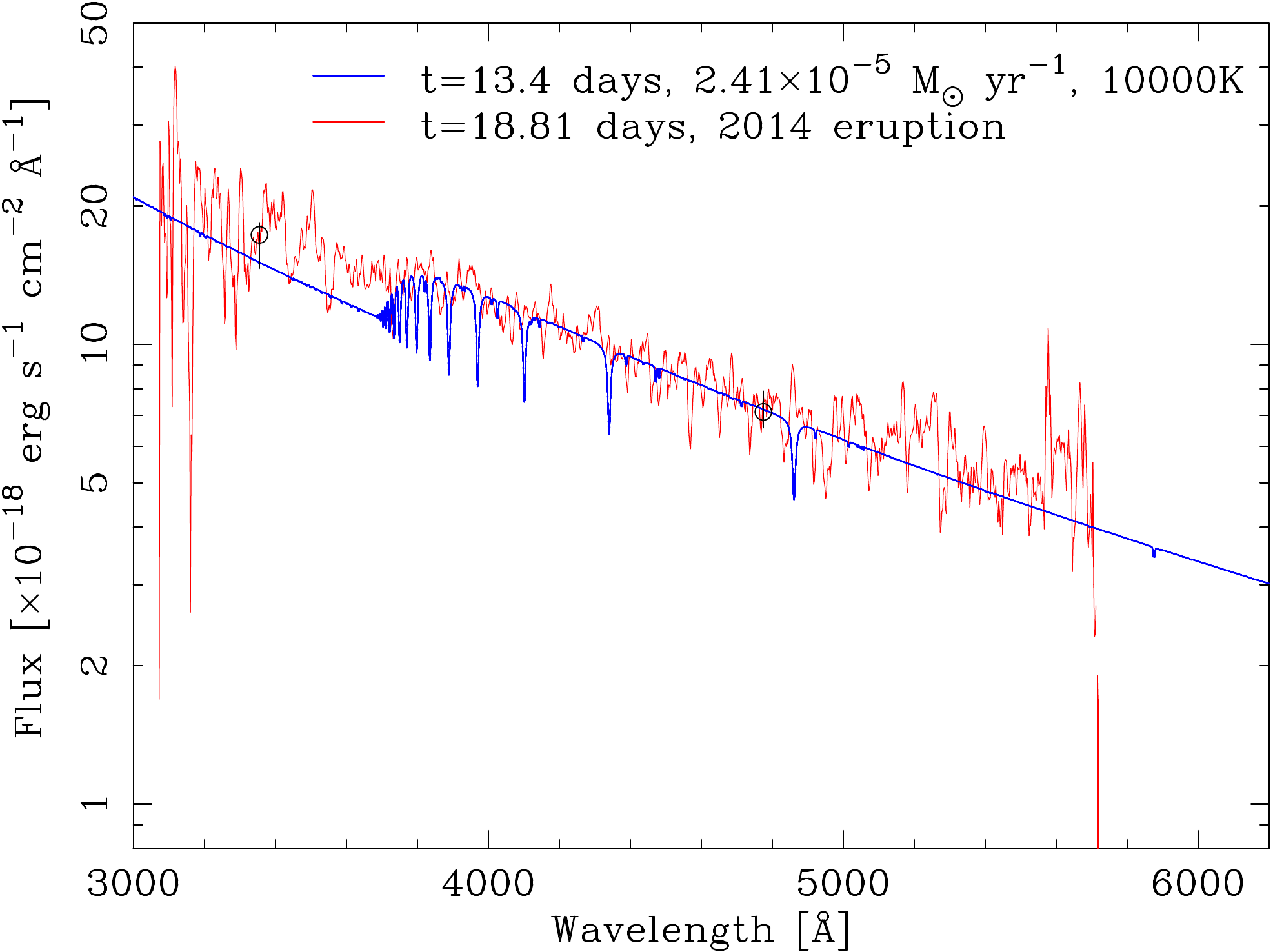}
\caption{Direct comparison between the $t=13.4$\,day accretion disk model of the 2015 eruption and the dereddened $t=18.81$\,day Keck spectrum of the 2014 eruption.  The flux of the Keck spectrum has been increased by $8.85\times10^{-18}$\,erg\,s$^{-1}$\,cm$^{-2}$\,\AA$^{-1}$.  The spectrum and the model are in good agreement above 4000\,\AA, however typical accretion disk absorption lines and the Balmer discontinuity is not present the spectrum, which contains H$\beta$ in emission.\label{keck_vs_model}}
\end{figure}

\begin{figure*}
\includegraphics[width=\columnwidth]{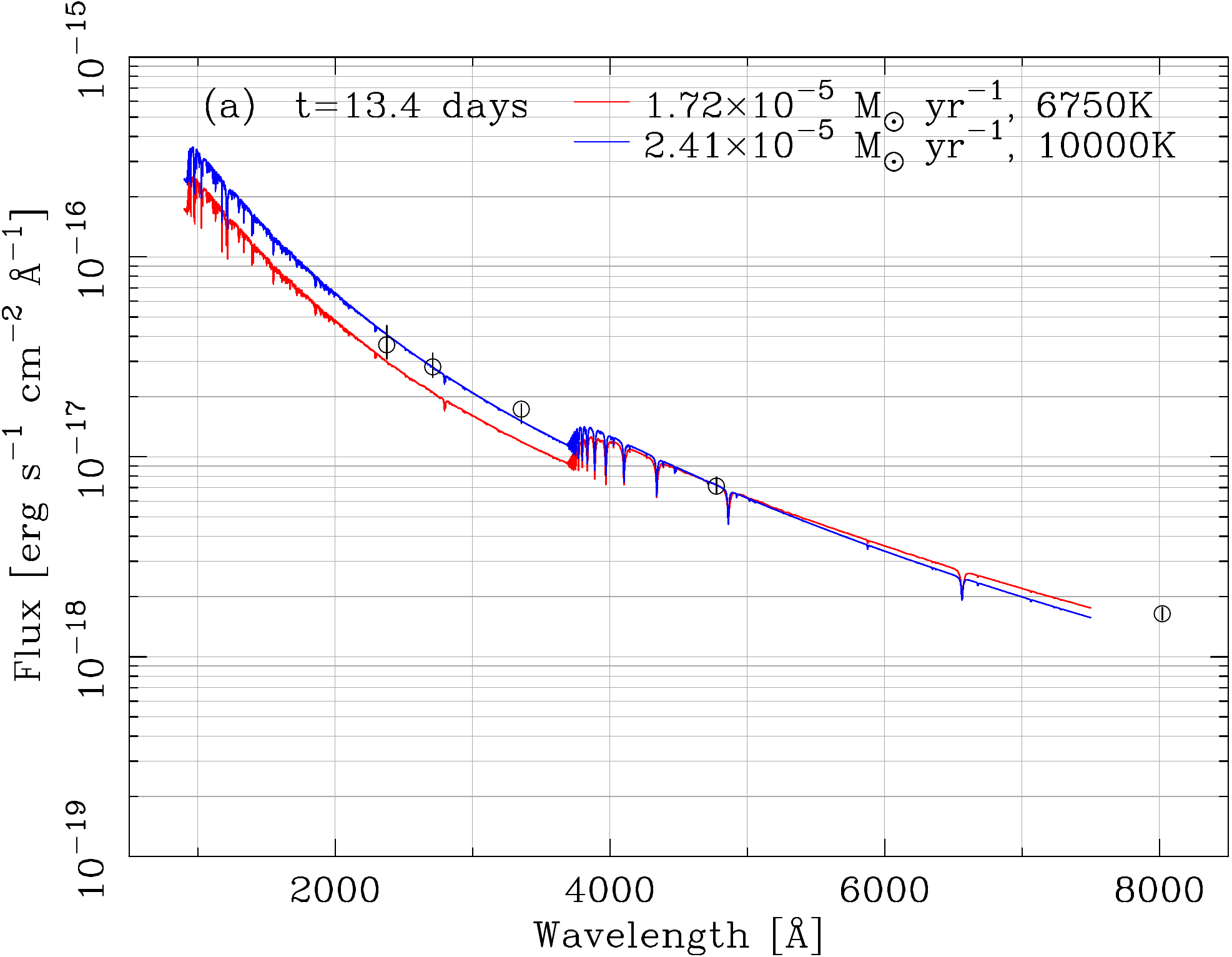}\hfill
\includegraphics[width=\columnwidth]{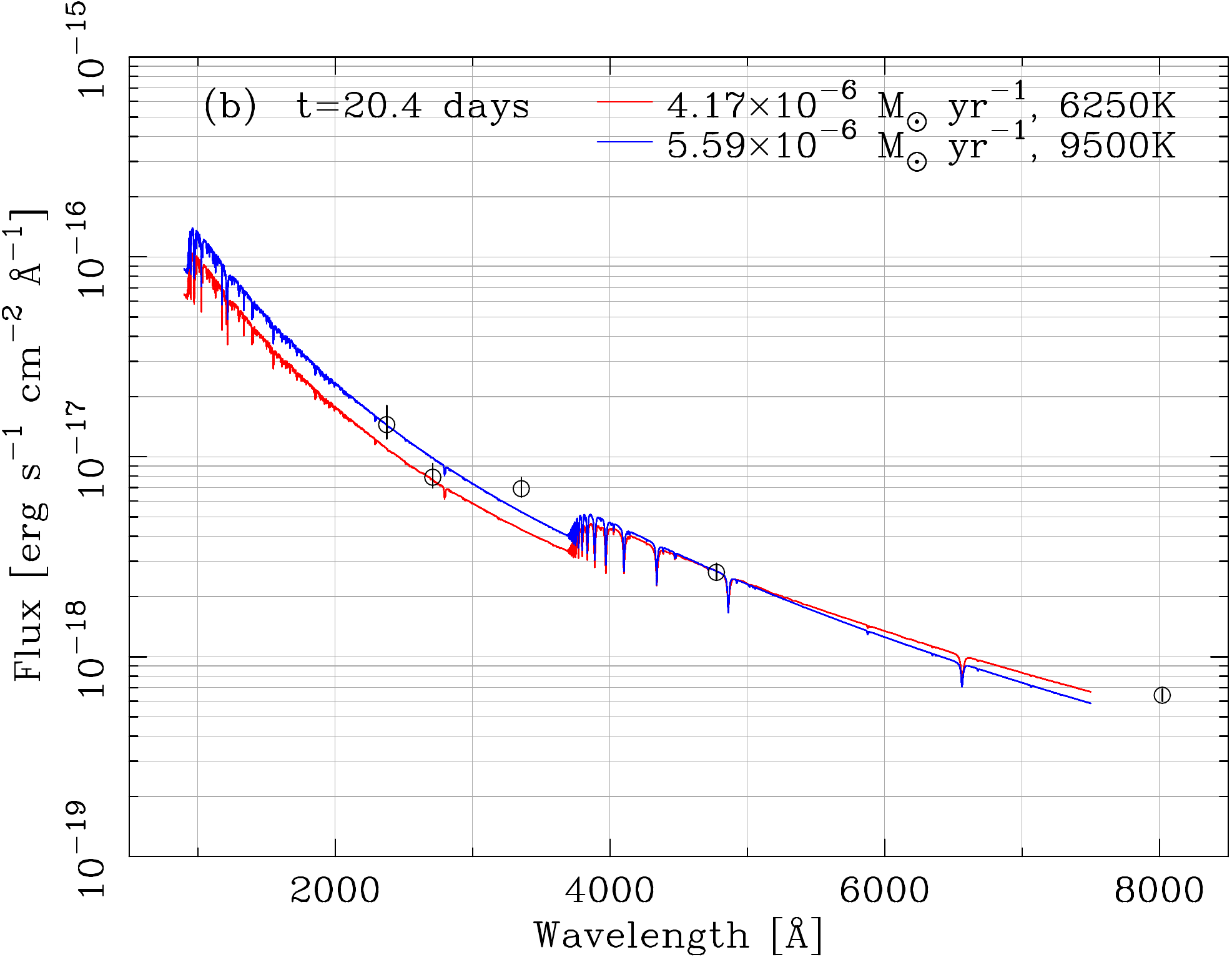}\\
\includegraphics[width=\columnwidth]{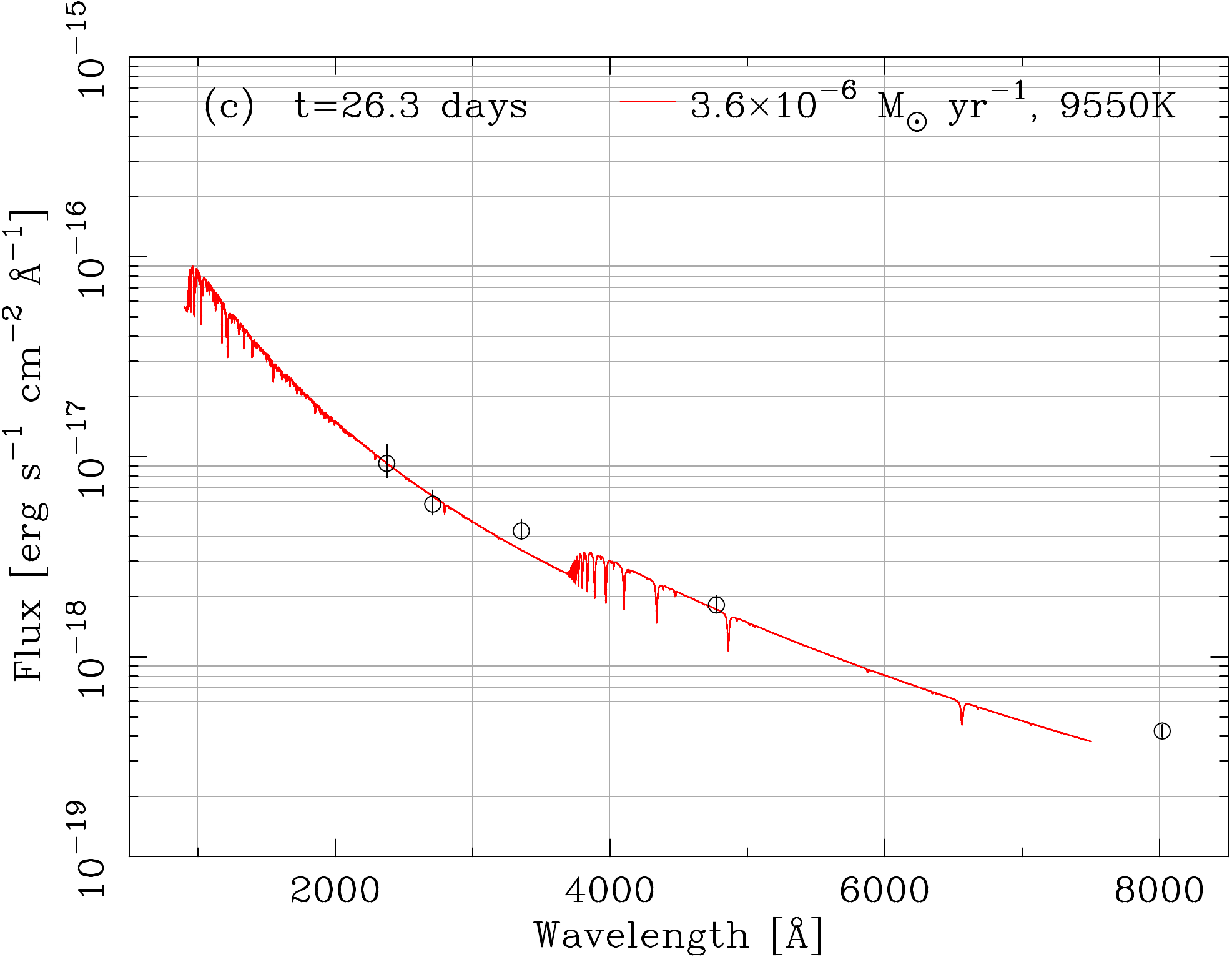}\hfill
\includegraphics[width=\columnwidth]{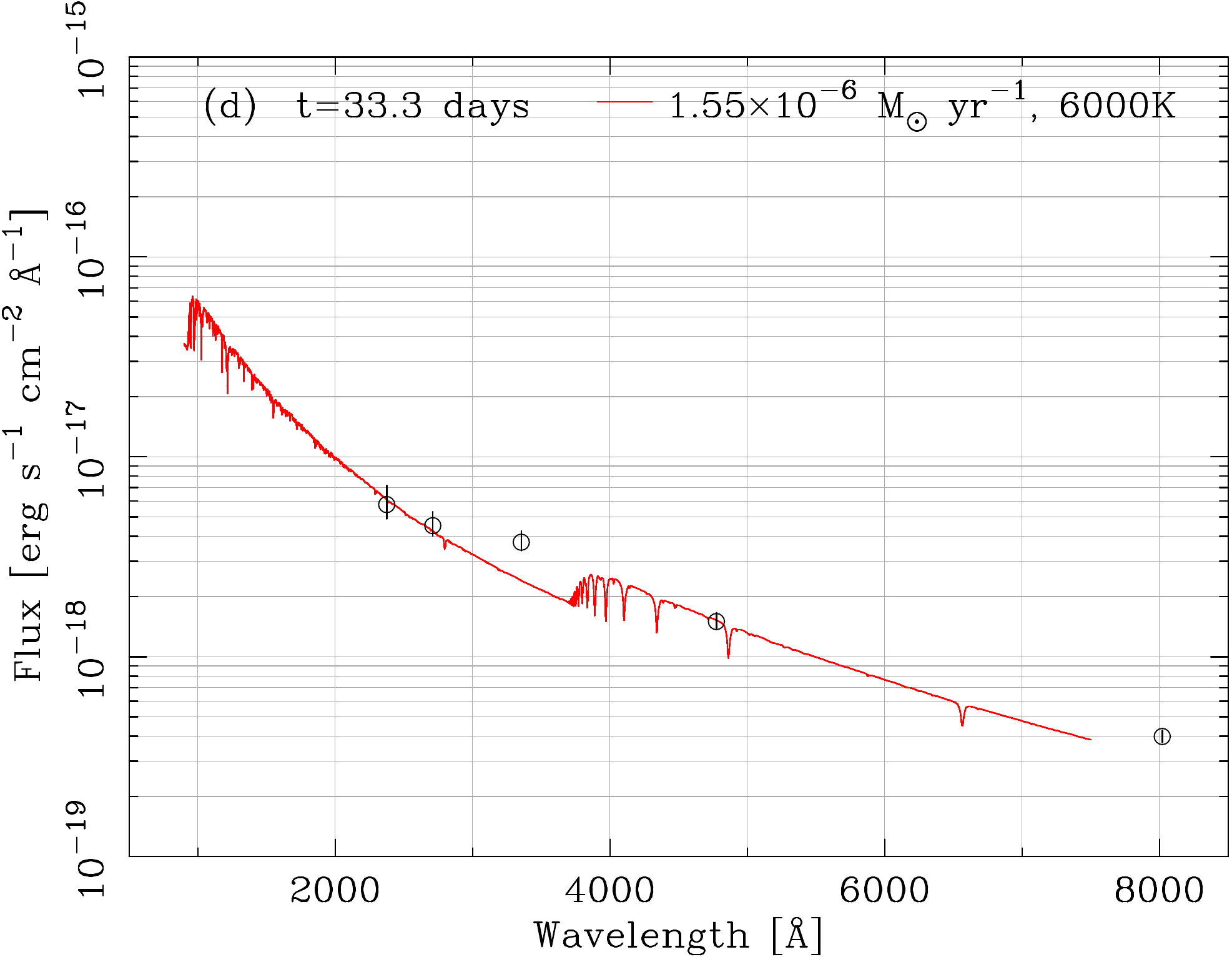}\\
\includegraphics[width=\columnwidth]{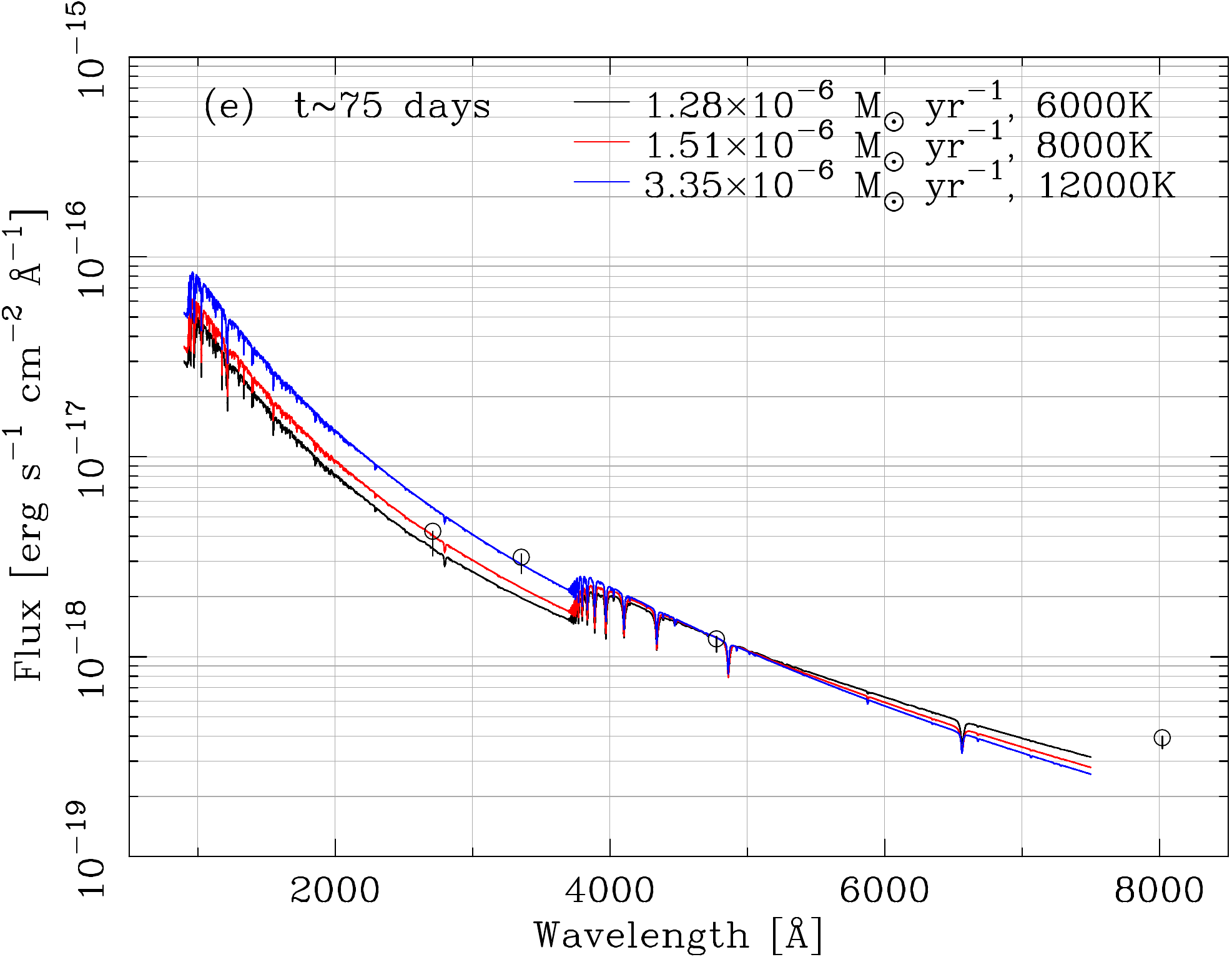}\hfill
\includegraphics[width=\columnwidth]{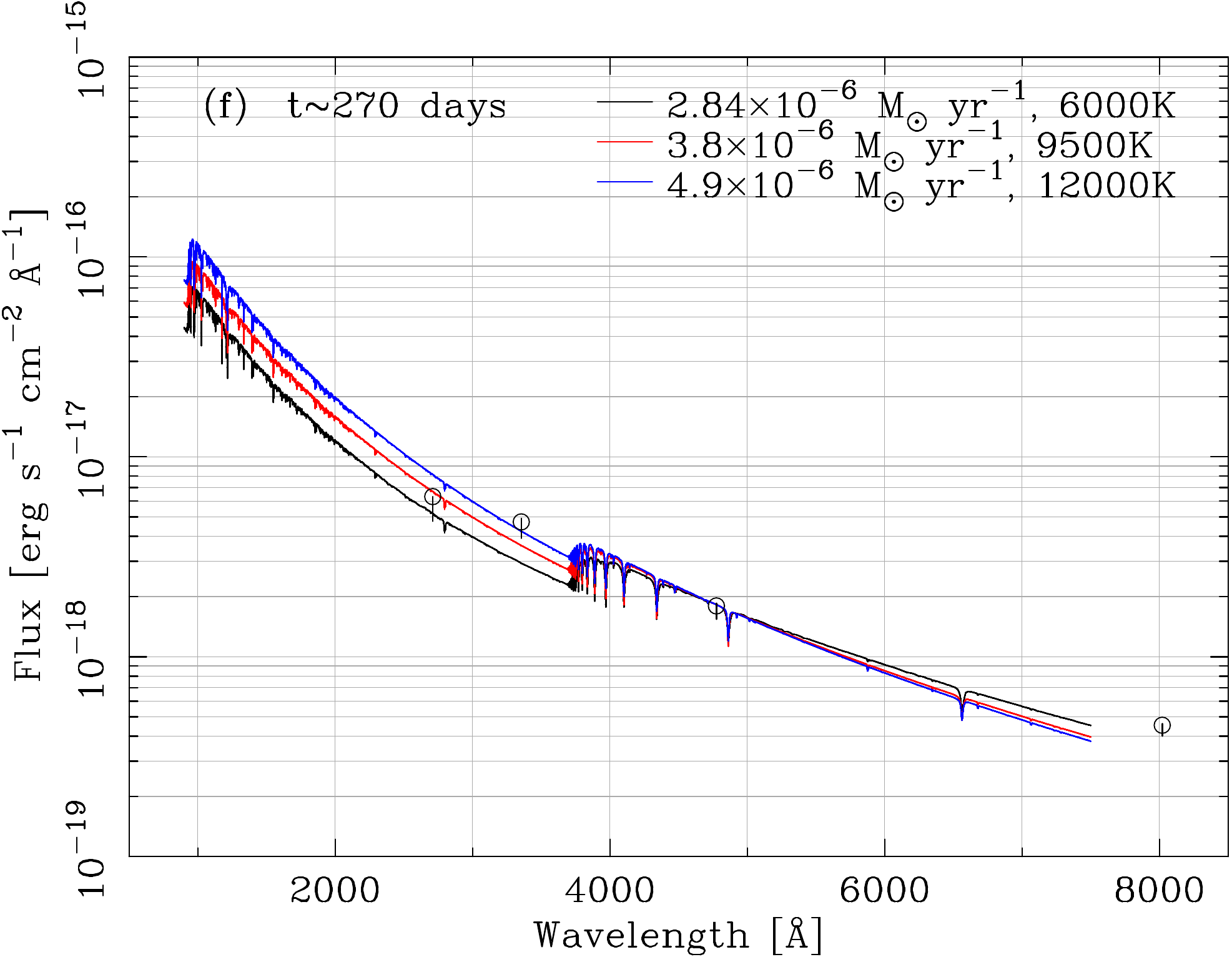}
\caption{Accretion disk modeling of the declining \novak.\label{mod:q}\label{mod:d}.  Each sub-plot indicates the epoch post-eruption, and the black data points show the corresponding broad-band {\it HST} photometry and associated uncertainty.  For model fit details, refer to the text and the keys in each sub-plot, which indicate the mass accretion rate $\dot{M}$ and the temperature at the truncated outer edge of the disk.  Sub-plots (a)--(d) cover the late decline of the 2015 eruption, (e) and (f) are the two reconstructed inter-eruption, or quiescent, epochs.}
\end{figure*}

For the disk models to simply provide sufficient flux to match the observations at the distance of M\,31, the disk mass accretion rate\footnote{We again note that this may be formally different from the WD mass accretion rate, $\dot{M}_\mathrm{acc}\leq\dot{M}$.}, $\dot{M}$, is required to be large, $\gtrsim10^{-6}\,M_{\odot}$\,yr$^{-1}$, and therefore not far away from $\dot{M}_\mathrm{Edd}$.  Such models generate a prominent Balmer discrepancy at $\sim4000$\,\AA, which is not apparent in the quiescent SEDs (also see the late-time spectrum in Figure~\ref{keck_vs_model}).  However, many CVs {\it accreting at a high rate} do not exhibit strong Balmer discontinuities \citep{2015MNRAS.450.3331M}.  We began by fitting the synthetic spectra longward of the Balmer discontinuity to just the F475W photometry (4773.7\,\AA ), which requires $\dot{M}=1.28 \times 10^{-6}\,M_{\odot}$\,yr$^{-1}$, see Figure~\ref{mod:q}e (solid black line).  This model has an outer disk radius extending to $1320\,R_\mathrm{WD}\simeq3.8\,R_{\odot}$, where the temperature falls to 6000\,K, but the model is clearly deficient in flux at wavelengths shorter than the Balmer discontinuity. 

\citet{2015MNRAS.450.3331M} proposed that the absence or reduction of the Balmer discontinuity observed in some CVs is due to continuum emission from a disk wind. We note that disk models whose outer radii are truncated also produce spectra with decreased Balmer discontinuities.  In Section~\ref{orb_per} we briefly discuss possible physical explanations for disk truncation.  Truncated disk models provide a slightly lower continuum flux level for the same $\dot{M}$. 

If the disk is truncated at $R=880\,R_{\rm wd}$ ($\approx 2.53\,R_{\odot}$), the model fits the first data point (shortest wavelength).  This effectively removes regions of the disk cooler than 8000\,K.  Outer disk truncation results in a reduced flux (for a fixed $\dot{M}$), therefore the mass accretion rate of this model must rise to $1.51\times10^{-6}\,M_{\odot}$\,yr$^{-1}$, see Figure~\ref{mod:q}e (red line).  To fit the second data point, nearest to the Balmer edge, we further truncate the disk to 750$\,R_{\rm WD}$ ($\approx 2.08\,R_{\odot}$). Such a disk has $\dot{M}=3.35 \times 10^{-6}\,M_{\odot}$\,yr$^{-1}$ and the temperature in the outer disk reaches 12000\,K. This model, however, overshoots the first  data point.  The model is shown in Figure~\ref{mod:q}e (blue line), and the derived accretion rates are tabulated in Tables~\ref{Mdot_Data}.   We note that all three models underestimate the F814W flux.  The excess flux here may be contributed by the donor (see Sections\,\ref{donor_excess} and \ref{sec:donor}).  We also note that the differing photometric points at both quiescent epochs were taken at different times, therefore any fundamental variability at quiescence could be imprinted on these data.

\begin{table}
\caption{Computed accretion rates for \novak\ during the final decline of the 2015 eruption and at quiescence.\label{Mdot_Data}}
\begin{center}
\begin{tabular}{llll}
\hline
Epoch & $\dot{M}$ & $\dot{M}_\mathrm{wind}\simeq\dot{M}_\mathrm{acc}$ & $\dot{M}_\mathrm{truncated}$ \\
(days) & \multicolumn{3}{c}{($\times 10^{-6}\,\mathrm{M}_\odot\,\mathrm{yr}^{-1}$)} \\
\hline
13.4 & 17.2 & 8.60 & 24.1\\
20.4 & 4.17 & 2.86 & 5.59\\
26.3 & 3.60 & 1.80 & 4.30\\
33.3 & 1.55 & 0.77 & 1.86\\
$\sim$75 & 1.28 & 0.64 & 3.35\\
$\sim$270 & 2.84 & 1.42 & 4.90\\
\hline
\end{tabular}
\end{center}
\tablecomments{$\dot{M}$ refers to the disk mass accretion rates as directly calculated by the accretion disk models, $\dot{M}_\mathrm{truncated}$ refers to the $\dot{M}$ {\it upper} limits imposed by heavily truncated disk, $\dot{M}_\mathrm{wind}$ is the expected maximum mass loss from the disk via a disk wind, with the remaining amount $\dot{M}_\mathrm{acc}$ being accreted onto the WD (i.e., $\dot{M}_\mathrm{acc}=\dot{M}-\dot{M}_\mathrm{wind}$, or $\dot{M}_\mathrm{acc,min}\simeq\dot{M}/2$; also see the later discussion about possible outflows).}
\end{table}

Next, we turn to the second quiescence epoch, $\sim270$\,d post-eruption, and (assuming an annual cycle) $\sim70$\,d pre-eruption.  These data are similar to those at $t\simeq75$\,d (see Figure~\ref{mod:q}e and \ref{mod:q}f), but the flux is higher.  Consequently,  we follow the same modeling procedure.  In Figure~\ref{mod:q}f we present three models with the outer disk truncated, and $(2.84\leq\dot{M}\leq4.9)\times 10^{-6}\,\mathrm{M}_\odot\,\mathrm{yr}^{-1}$.  Again, truncating the cooler outer disk reduces the ``jump'' of the Balmer edge. From $t\simeq75$\,d to $t\simeq270$\,d $\dot{M}$ has increased by a factor of $\sim 1.5$ to 2.2.  

In all the models presented here, we found that the inclusion of a hot WD did not contribute any significant flux due to the small surface area of the massive WD and to the very large area of the very hot disk. It is also probable that at high $\dot{M}$ the inner disk is swollen and masks the WD.

\subsubsection{The decline} 

We next consider the evolution during the late decline of the 2015 eruption.  We model these in reverse, as the complexity of the emission is expected to increase closer to the eruption itself. 

At $t=33.3$\,d, the flux from \novak\ lies approximately midway between that at quiescence ($t\simeq75$ and $\simeq270$\,d, see left panel of Figure~\ref{fig:sed_evo}), and we find that a standard (non-truncated) disk model with $\dot{M}=1.55\times10^{-6}\,\mathrm{M}_\odot\,\mathrm{yr}^{-1}$ provides a reasonable fit to the data. This model has an outer region extending to where the temperature reaches 6000\,K, extending the outer region to 3500\,K does not improve the fit to the data points.  The fit is presented in Figure~\ref{mod:d}d.  There is a slight flux excess at $\sim$3350\,\AA, but as shorter wavelengths are consistent with the model, a truncated disk model does not provide a better fit to the data.  Again, there is a flux excess at $\sim$8000\,\AA.

Turning to $t=26.3$\,d, the data points are in better agreement with the presence of a weak Balmer edge (see Figure~\ref{mod:d}c). We fit a disk model while varying the outer truncation radius, and find that the best fit is obtained for $\dot{M}\sim 3.6 \times 10^{-6}\,\mathrm{M}_\odot\,\mathrm{yr}^{-1}$ with the outer disk truncated at 9500\,K ($R_\mathrm{disk}=1050\,R_\mathrm{WD}$, $\sim 3\,R_{\odot}$). 
 
A week earlier, $t=20.4$\,d, we find $\dot{M}=4.17\times10^{-6}\,\mathrm{M}_\odot\,\mathrm{yr}^{-1}$ and $5.59\times10^{-6}\,\mathrm{M}_\odot\,\mathrm{yr}^{-1}$ for disks truncated at 6250\,K and 9500\,K, respectively.  Neither model reproduces the NUV flux well, possibly an indication of a contribution from an additional source or lines.  These two disk models are presented in Figure~\ref{mod:d}b.    

\subsubsection{The Super-Soft X-ray phase.} 

Finally we turn to the observations at $t=13.4$\,d, during the SSS phase of the 2015 eruption. As is evident from Figure~\ref{mod:d}a, not only is this the epoch with the highest flux, but the data exhibit a rather smooth ``continuum'' -- almost a straight line on this logarithmic scale.  This is further illustrated by the Keck spectrum taken 18.81\,d after the 2014 eruption, which is directly compared to the 2015 $t=13.4$\,d data in Figure~\ref{keck_vs_model}.

There is no indication of the presence of the Balmer edge from the 2015 data, which is confirmed by the Keck 2014 spectrum.  A disk model truncated at 20000\,K produces a smooth continuum without a Balmer edge, but the continuum slope is much steeper than inferred from the data.  Therefore, we fit the data with disk models that have various degrees of truncation. We find $\dot{M}=1.72 \times 10^{-5}\,\mathrm{M}_\odot\,\mathrm{yr}^{-1}$ for a disk truncated at 6750\,K, and $2.41 \times 10^{-5}\,\mathrm{M}_\odot\,\mathrm{yr}^{-1}$ for a disk truncated at 10000\,K; see Figure~\ref{mod:d}a.  As with the other epochs, these models cannot fit all the data points simultaneously.  We note that such an $\dot{M}$ is above $\dot{M}_\mathrm{Edd}$.

\subsection{Disk winds}

As mentioned earlier, \citet{2015MNRAS.450.3331M} proposed that the absence or reduction of the Balmer edge in optical spectra of some CVs is due to the existence of powerful accretion disk winds. Indeed, \citet{2015MNRAS.450.3331M} show that a standard disk wind model is successful in reproducing the weak Balmer absorption edge at all inclinations, but particularly for CV systems viewed at high inclination. They further suggest that winds can dominate the continuum emission from CVs. Their modeling shows that the inclusion of the disk wind produces a much weaker Balmer edge, a shallower continuum slope, and the flux level increases due to the contribution of the wind to the disk continuum. 

Consequently, compared to the model fit in \citet{2015MNRAS.450.3331M}, our optically thick non-truncated standard disk models provide an {\it upper limit} to the mass accretion rate onto the WD ($\dot{M}_\mathrm{acc}$), as they produce less flux at the same accretion rate. The discrepancy between the wind disk model and the standard disk model is minimal near the upper edge of the Balmer jump ($\sim 4000-5000$\,\AA ) and appears to reach a maximum of about a factor of two in $\dot{M}$. Therefore, if we assume that disk wind emission has to be taken into account, we have to reduce the mass accretion rates obtained from our non-truncated disk model fits by a maximum of 50\% (i.e., $\dot{M}_\mathrm{acc}\simeq\dot{M}_\mathrm{wind}\simeq0.5\dot{M}$).  Subsequently, the implied mass accretion rates in the presence of a disk wind are shown in Table~\ref{Mdot_Data}.

\subsection{Model uncertainties}\label{model_uncert}
 
Finally, we compute the relative uncertainties introduced from the errors on the reddening, inclination, distance, and  fluxes.  For this purpose we consider the data for $t=33.3$\,d, with $\dot{M}=1.55\times10^{-6}\,\mathrm{M}_\odot\,\mathrm{yr}^{-1}$.  

\hstk\ computed that the reddening toward \novak\ is $E_{B-V}=0.10\pm0.03$.  De-reddening the $t=33.3$\,d data using $E_{B-V}=0.07$ and $E_{B-V}=0.13$ gives $\dot{M}=1.33\times10^{-6}\,\mathrm{M}_\odot\,\mathrm{yr}^{-1}$ and $\dot{M}=1.90\times10^{-6}\,\mathrm{M}_\odot\,\mathrm{yr}^{-1}$, respectively.  That is, the disk mass accretion rate becomes $\dot{M}= 1.55^{+0.35}_{-0.22} \times 10^{-6}\,\mathrm{M}_\odot\,\mathrm{yr}^{-1}$.

Similarly we compute the errors for an inclination of  $i=20^{\circ} \pm 10^{\circ}$, distance of 770$\pm 20$ kpc, and a maximum error of 5\% in the fluxes (see Table~\ref{PHAT_phot}).  Assuming $\dot{M}$ varies linearly with small changes in $E_{B-V}$, $i$, $d$, and the fluxes, by quadrature we obtain $\dot{M}= 1.55 ^{+0.39}_{-0.25} \times 10^{-6}\,\mathrm{M}_\odot\,\mathrm{yr}^{-1}$, errors of $+25\%$ and $-16\%$.  These errors are much smaller than the systematics introduced from the use of a truncated disk model (a factor of $\sim2$ in $\dot{M}$) or when comparing our standard disk models to the \citet{2015MNRAS.450.3331M} disk wind models (a factor of $\sim0.5$).  Namely, in fitting the data from $t=13.4$\,d, we obtained $\dot{M}=1.28 \times 10^{-6}\,\mathrm{M}_\odot\,\mathrm{yr}^{-1}$, but $\dot{M}$ could be about twice this value if we consider the truncated disk models, or could be about half this value if we consider the possibility of a disk wind continuum.  

\begin{figure}
\includegraphics[width=\columnwidth]{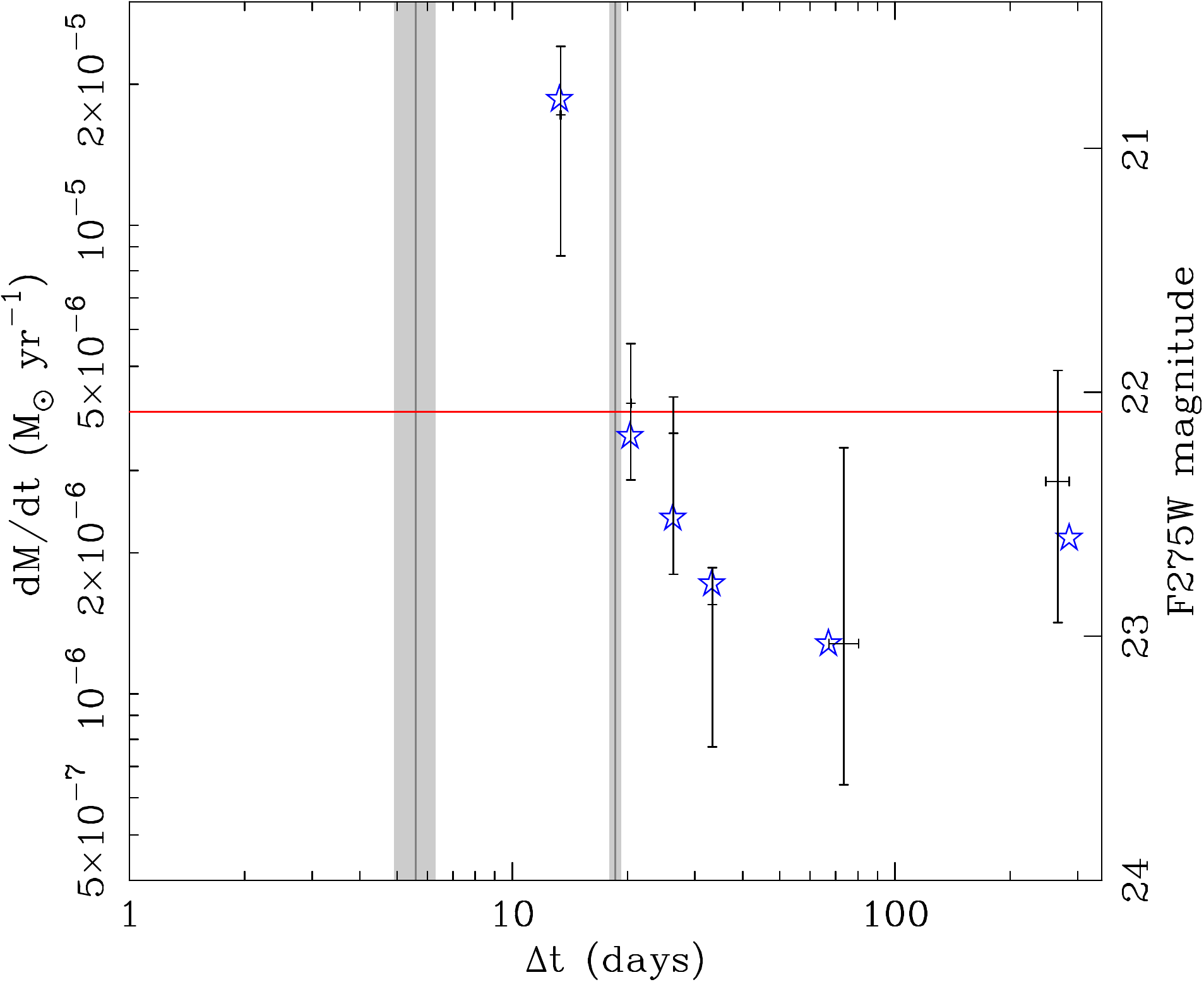}
\caption{Evolution of the disk mass accretion rate ($\dot{M}$) of the \novak\ accretion disk with time within an eruption cycle (black points).  The upper limits refer to $\dot{M}_\mathrm{truncated}$ and the lower limits $\dot{M}_\mathrm{wind}(\simeq\dot{M}_\mathrm{acc})$, see Table~\ref{Mdot_Data}.  The {\it HST} F275W photometry is also plotted for comparison (blue data points), and to indicate that it traces the accretion rate well.  The Eddington accretion limit of a 1.37\,M$_\odot$ WD is indicated by the red horizontal line.  The vertical gray lines indicate the turn on and turn off times of the SSS from the 2015 eruption (the shaded areas their associated uncertainties).\label{mod:t}}
\end{figure}

Taking this into account, we reproduce the computed accretion rates in Table~\ref{Mdot_Data}, and plot them as a function of time in Figure~\ref{mod:t}. 

\subsection{Donor flux excess}\label{donor_excess}

As can be seen in Figure~\ref{mod:d}, there is a flux excess, above the disk models, in the F814W band at all epochs.  As the only expected `red' component in the system, this excess flux is probably from the donor.  To examine this effect, we extended the $t\simeq75$\,d and $t\simeq270$\,d accretion disk models to longer wavelengths by fitting a power law to the model spectra, redward of the Balmer limit.  This produced a good fit to the model spectra, and enabled us to determine a F814W flux excess at quiescence of $(1.67\pm0.18)\times10^{-19}$\,erg\,s$^{-1}$\,cm$^{-2}$\,\AA$^{-1}$, which corresponds to an apparent magnitude of $m_\mathrm{F814W}=24.8_{-0.1}^{+0.2}$.  Further extrapolation of the accretion disk model confirms that any disk contribution in the NIR F110W and F160W filters is negligible. 

\section{Discussion}\label{sec:disc}

\subsection{The accretion disk}

In Section~\ref{sec:mod} we described the comparison between the {\it HST} photometry of the final decline of the 2015 eruption, and quiescent observations, of \novak\ to models of accretion disks around 1.37\,M$_\odot$ WDs.  We again state that the ideal datasets for such work would be spectroscopy extending into (and even beyond) the FUV.  However, for CVs at the distance of M\,31 such observations are not yet feasible.  Therefore, the {\it HST} visible and NUV photometry described in this paper currently provide the best, {\it and only}, data with which to explore the accretion disk in \novak.

The one thing that is immediately clear is the very large luminosity of the \novak\ accretion disk at quiescence. By necessity, modeling of such a high luminosity disk requires a large disk mass accretion rate ($\dot{M}$).  The results of the modeling show that the broad-band photometric SED of \novak\ from the epoch of the first post-eruption {\it HST} imaging, and during quiescence, is consistent with the expected form of an accretion disk.  As the first {\it HST} epoch occurs only 13\,days after the 2015 eruption, indeed before the SSS is extinguished, this is evidence that the accretion disk may survive eruptions of \novak.

The basic form of the SED, from the optical to NUV, remains consistent from $t=13$\,d to quiescence, adding further weight to the survival of the disk.  Observationally, we first see this disk beginning to dominate the optical/NUV emission about two weeks post-eruption, and possibly as early as $t=4$\,days\footnote{Could the optical/NUV light curve plateau presented in \othreek\ be caused by the surviving disk being unveiled by the receding photosphere?  A similar prediction was made for a number of Galactic RNe by \citet{2008ASPC..401..206H}.}.  The disk luminosity decreases to a minimum just $\sim75$\,days post eruption, before building again toward the next eruption -- presumably as the accretion disk increases in mass.  Our working model is that the disk, once struck by the nova ejecta, is initially shocked and heated, but survives largely intact.  Further, irradiation from the SSS may begin to affect the disk from as early as $t=4$\,days.  These effects cause the disk to begin losing mass at a large rate through a disk wind (with the disk accretion rate at $\dot{M}\sim2\times10^{-5}\,M_\odot\,\mathrm{yr}^{-1}$), as is discussed below, some of this mass may be accreted directly onto the WD.  As the surviving disk then cools and relaxes its luminosity decreases until reaching a minimum after $\sim75$\,days ($\dot{M}\sim10^{-6}\,M_\odot\,\mathrm{yr}^{-1}$).  During the next $\sim200$\,days of quiescence, mass loss from the donor allows the disk to rebuild any matter lost (through the eruption and disk wind), in the run up to the next eruption.

But there is a potential problem, not necessarily with the picture outlined above, but with the mass accretion rates derived from the models, which do not include disk winds.  Namely, we computed values of $\dot{M}$ representing the disk mass accretion rate, {\it not} the accretion rate onto the WD ($\dot{M}_\mathrm{acc}$).  Up to half of $\dot{M}$ might be lost through a disk wind \citep{2015MNRAS.450.3331M}, reducing the effective accretion onto the WD by up to 50\%. Our disk models imply that $\dot{M}$ is close to, or even exceeds, $\dot{M}_\mathrm{Edd}$ throughout the entire eruption cycle, a state where a significant radiation pressure driven disk wind may be expected to be present. 

A number of authors have investigated the WD mass -- WD accretion rate ($\dot{M}_\mathrm{acc}$) phase space, and they arrive at two broad but differing conclusions.  The first is that, other than $\dot{M}_\mathrm{Edd}$ itself, there is (for a given WD mass) no upper limit on the mass accretion rate, and that nova eruptions will occur at any $\dot{M}_\mathrm{acc}$ \citep[see, e.g.,][]{Starrfield2016}.  Or alternatively, that there is a clear upper limit to $\dot{M}_\mathrm{acc}$ \citep[see, e.g.,][]{1982ApJ...257..752F,1982ApJ...253..798N}, beyond which nova eruptions cease, with the WD entering a phase of steady state nuclear burning (the persistent SSS).  At even higher accretion rates, these models predict that optically thick winds (from the WD) are generated.  In recent years, it has been proposed that one important difference between these two scenarios is how mass accretion is treated during a nova eruption \citep{2016ApJ...824...22H}, with the former assuming it ceases, the latter assuming it continues.  With \novak\ showing both signs of a surviving disk, therefore continuing accretion, and an elevated $\dot{M}$, it may be an important system in addressing this long-standing issue.

\begin{figure}
\includegraphics[width=\columnwidth]{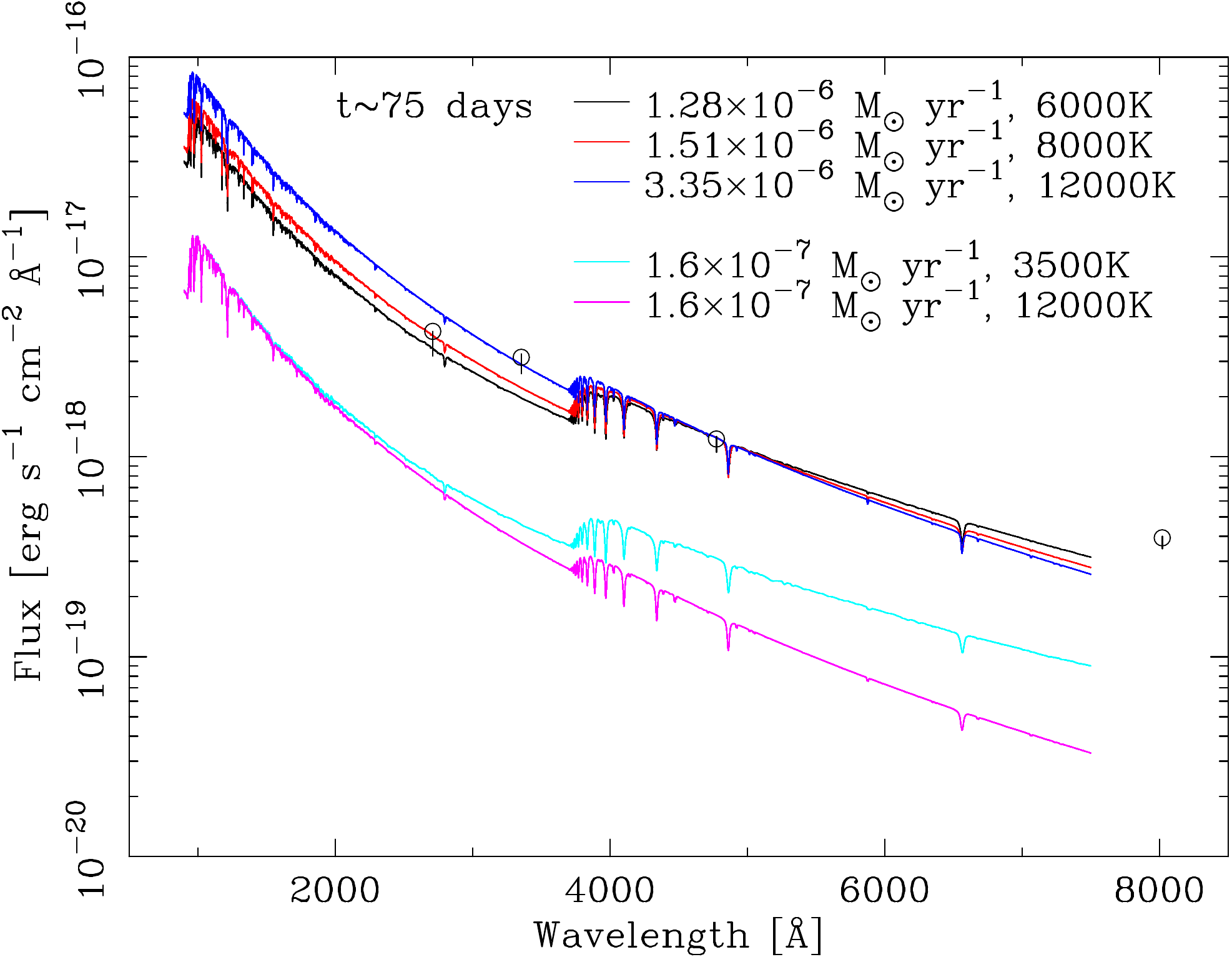}
\caption{As Figure~\ref{mod:q}, comparing the best fit accretion disk models to the {\it HST} photometry $\sim75$\,days post-eruption -- the minimum state.  Also shown, in magenta and cyan, are our predicted accretion disk spectra based on an accretion rate of $\dot{M}=1.6\times10^{-7}\,M_\odot\,\mathrm{yr}^{-1}$, as required by \citet{2014ApJ...793..136K,2015ApJ...808...52K,2016ApJ...830...40K,2017ApJ...838..153K}.  Disks with the \citeauthor{2015ApJ...808...52K} accretion rate under-predict the quiescent flux of \novak\ by a factor of $\sim10$.\label{newvsold}}
\end{figure}

Discussion of the merits of these two differing pictures is clearly beyond the scope of this paper.   But we note, of course, that the former (with no upper limit) poses no clear obstacle to our derived accretion rates.  Turning to the latter, we note that the work of \citet{2014ApJ...793..136K,2015ApJ...808...52K,2016ApJ...830...40K,2017ApJ...838..153K,2017ApJ...844..143K}, employing such a formulation, has already successfully modeled many observational aspects of the \novak\ eruptions, while assuming a {\it constant} $\dot{M}_\mathrm{acc}=1.6\times10^{-7}\,M_\odot\,\mathrm{yr}^{-1}$ -- a factor of four lower than the $\dot{M}_\mathrm{acc}$ lower limit derived in this work (under the assumption of $\dot{M}_\mathrm{acc}\simeq\dot{M}_\mathrm{wind}$).  As is shown graphically in Figure~\ref{newvsold}, accretion disks with $\dot{M}=1.6\times10^{-7}\,M_\odot\,\mathrm{yr}^{-1}$ significantly under-predict the NUV flux of \novak\ at quiescence.  The discrepancy is a factor of $\sim10$, even at the quiescence minimum of $\sim75$\,days post-eruption.

\begin{figure}
\includegraphics[width=\columnwidth]{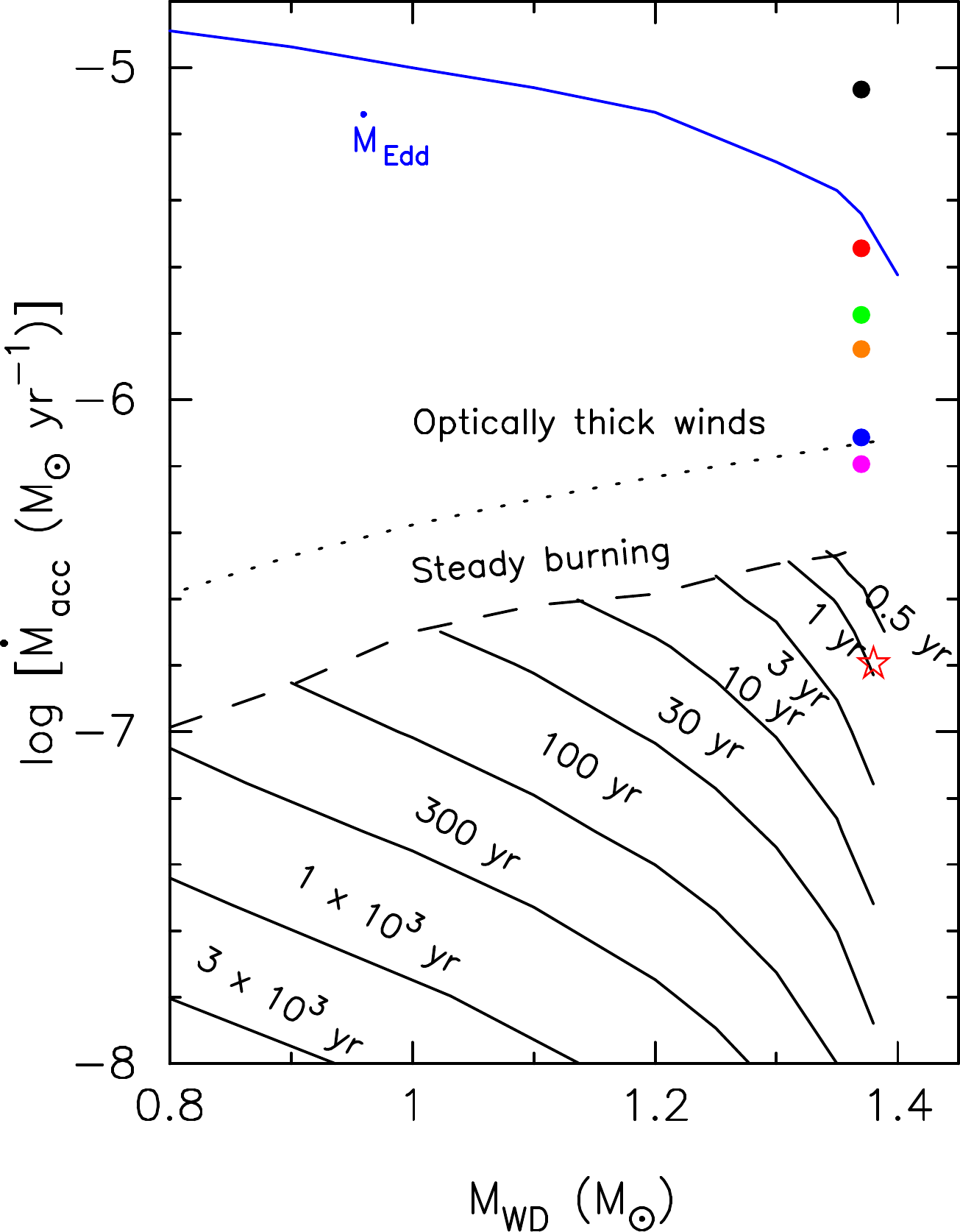}
\caption{Recurrence period $t_\mathrm{rec}$ of novae on the $M_\mathrm{WD}-\dot{M}_\mathrm{acc}$ plane, based on Figure~6 of \citet{2014ApJ...793..136K}. We plot the loci of the equi-recurrence periods of novae (black solid lines). Hydrogen burning is proposed to be stable in the region above the dashed line ($\dot{M}_\mathrm{stable}$). In the region below $\dot{M}_\mathrm{stable}$, H-shell burning is thermally unstable, and the WD experiences shell flashes (i.e., novae). Optically thick winds are accelerated in the region above the dotted line ($\dot{M}_\mathrm{cr}$).  We note that the \citet{2016PASP..128e1001S} interpretation of this plot does not require either the steady burning nor optically thick wind regions (see text for more details).  The solid blue line indicates the Eddington limit \citep[based on][]{1982ApJ...253..798N}.  The red star indicates the quiescent position of \novak, based on the modelling in \citet{2015ApJ...808...52K}.  The coloured points, at $M_\mathrm{WD}=1.37\,M_\odot$ indicate the {\it lower limits} of the computed values of $\dot{M}$ (i.e.\ assuming that  $\dot{M}_\mathrm{acc}=0.5\dot{M}$) from (top--bottom) t=13.4 (black), 20.4 (red), 26.3 (green), $\sim$270 (orange), 33.3 (blue), $\sim$75\,d (magenta) post-eruption (see $\dot{M}_\mathrm{wind}\simeq\dot{M}_\mathrm{acc}$ values in Table~\ref{Mdot_Data}).\label{kato_plot}}
\end{figure}

In Figure~\ref{kato_plot} we have recreated Figure~6 of \citet[M.\ Kato, priv.\ comm.]{2014ApJ...793..136K}, which shows the loci of equi-recurrence periods of novae in the WD mass -- $\dot{M}_\mathrm{acc}$ plane.  In this plot, as discussed above, the regions of proposed steady burning and optically thick winds are shown.  The position of \novak\ as computed by \citet{2014ApJ...793..136K} is indicated by the red star, and this lies clearly at the extremes of the phase-space permitted by these models.  The disk mass accretion rates computed in this work are clearly at odds with the \citet{2014ApJ...793..136K} formulation, unless only a small proportion of the matter from the disk is accumulated on the WD surface.  Given our computed mass loss rates, we would require {\it at least} 80\% of $\dot{M}$ to constitute a disk wind ($\dot{M}_\mathrm{wind}$), to stop the system undergoing the proposed steady-state nuclear burning.  However, such an elevated $\dot{M}_\mathrm{wind}$ seems unlikely for a sub-critical accretion disk \citep[see, e.g.,][]{2007MNRAS.377.1187P}.

Since our disk model indicates such a large $\dot{M}$, we must explore the validity of the standard $\alpha$/\citet{1973A&A....24..337S} disk model as $\dot{M}$ approaches $\dot{M}_\mathrm{Edd}$, since the basic ``geometrically thin'' assumption breaks down, i.e.\ height/radius $\approx 1$.  In this regime of $\dot{M} \approx \dot{M}_\mathrm{Edd}$, the disk can be represented using the slim-disk equations \citep{1988ApJ...332..646A}, where radial advection and radiation of energy is taken into account, and the flow can be partially supported by gas and radiation pressure. The departure from the standard disk model, however, is noticeable at mass accretion rates reaching $\dot{M} \sim 20 \dot{M}_{\rm Edd}$ \citep{1988ApJ...332..646A} as the heat trapped within the matter becomes important, and the luminosity increases more slowly than the accretion rate as the matter with its energy content is advected and radiated inward to the inner disk and onto the WD surface.  Since the maximum mass accretion we compute in this work is $\dot{M} \approx 4 \dot{M}_\mathrm{Edd} < 20 \dot{M}_\mathrm{Edd}$, and the minimum is as low as $0.2\dot{M}_\mathrm{Edd}$ during quiescence, our disk models are probably not strongly affected by neglecting advection of energy.

Advection of energy is, however, more pronounced in the inner disk and can be expected to peak in the boundary layer between the WD and disk, since an additional $L_{\rm bl} \approx L_{\rm acc}/2$ is released in that region.  This does not affect our disk models either, as the inner annuli in our models do not contribute significant flux at wavelengths $ > 2000$\,\AA, because of their small surface area ($r < 6.5 R_\mathrm{WD}$) and elevated temperature ($>3\times10^5$\,K) peaking in the EUV/soft X-ray regime ($\sim 10^5$\,K). 

Having established that our disk models are valid, we furthermore consider the fate of the advected energy in the inner disk/boundary layer. It has been shown that even at moderately large accretion rates ($\dot{M} \approx \dot{M}_{\rm Edd}$), advection of energy becomes important in the boundary layer \citep{1997ApJ...483..882G,1997ApJ...478..734P}.  As in advection dominated accretion flows \citep[ADAFs;][]{1994ApJ...428L..13N,1995ApJ...444..231N} the inner disk and boundary layer will radiate significantly less than expected, and the advected energy will heat up the WD and drive a bipolar outflow ($\dot{M}_\mathrm{bl}$) in addition to the disk-wind component. This will reduce the amount of material actually accreting onto the WD surface and could bring the WD accretion rate back toward the regime favoured by \citet[and others]{2015ApJ...808...52K}.  The outflow of matter is possibly low in the outer disk and increases inward, where a strong wind forms a bipolar outflow.  

For a number of decades some CVs have been suspected to have strong outflows with some systems even exhibiting ejecta, such as  the `nova-like' BZ\,Camelopardalis that is surrounded by a bow-shock nebula \citep{1984PASP...96..283E}.  However, so far, one finds no collimated outflows (``jets'') in CVs \citep{2004PASP..116..397H} in spite of the fact that all other disk systems (from X-ray binaries to AGN) exhibit collimated outflows \citep{1997ASPC..121..845L}. Is it possible that \novak\ is the exception to the {\it rule}, not just during eruption (see \othreek\ and \hstk) but at quiescence?

If we assume a strong disk wind, then about half of the disk material is accreted on to the WD (at a rate of $\dot{M}_\mathrm{acc}\sim6.4\times10^{-7}\,M_\odot\,\mathrm{yr}^{-1}$ at the apparent quiescent minimum), and the other half is deposited into the system (at the same rate, see Table~\ref{Mdot_Data} and Section~\ref{sec:donor}) -- the circumbinary environment. We note that the estimated red giant wind mass (total) in RS\,Oph at the time of eruption is $\sim10^{-6}\,M_\odot$ \citep{2011ApJ...740....5V}, broadly consistent with the circumbinary contamination predicted by the \novak\ disk wind.  Therefore, such a disk wind mass loss rate alone could be sufficient to account for the observed ejecta deceleration (\othreek) {\it without a requirement for a wind from the donor}.

The discussion of the accretion disk wouldn't be complete without considering irradiation of the outer disk by the hot inner disk/boundary layer region.  Disk irradiation is known to be important in low-mass X-ray binaries, where accretion occurs onto a neutron star or a black hole, with a much deeper gravitational potential well, while it is usually negligible in CVs \citep{1994A&A...290..133V,1998MNRAS.295L...1S,1998MNRAS.296L..45K}.  However, the WD in \novak\ is very compact with a mass of $1.37\,M_{\odot}$, a radius $R_\mathrm{WD}=2000$\,km and it is accreting at, or close to, the Eddington limit.  We therefore checked the importance of disk irradiation using the approach given by \citet{1990A&A...235..162V} for different values of $\dot{M}$.  At low disk mass accretion rates ($\dot{M} \lesssim 10^{-7}M_{\odot}$\,yr$^{-1}$; as typical for all other quiescent novae) irradiation increases the outer disk temperature by up to $\sim 1000$\,K, which does not produce any significant change in the disk spectrum. At more moderate accretion rates ($\dot{M}\sim 10^{-6} M_{\odot}$\,yr$^{-1}$; i.e., \novak\ at quiescence), irradiation increases the outer disk temperature by up to $\sim3000$\,K, thereby slightly affecting the spectrum by effectively decreasing the mass accretion rate, since irradiation increases the disk emission.  At mass accretion rates above the Eddington limit ($\dot{M} \sim 10^{-5}M_{\odot}$\,yr$^{-1}$; \novak\ during the eruption), we find that irradiation increases the outer disk temperature by as much as 7000\,K, and we would expect that this increase could reduce the mass accretion rate by a factor of $\sim2$. 
The effect of irradiation within \novak\ {\it at quiescence} is therefore only expected to slightly decrease the discrepancy in the disk mass accretion rate and the WD mass accretion rate. 

Returning finally to the survival of the accretion disk.  The presence of a disk, potentially with a high mass accretion rate, during the SSS phase of a nova eruption opens an intriguing possibility.  Could a surviving accretion disk continue to feed significant fuel to the nuclear burning region on the WD, thereby `artificially' extending the SSS phase, compared to a more typical nova (where the disk is assumed to be obliterated by the eruption)?  Such a `refuelling' would, for a short time, be akin to the persistent SSSs.  Any mass accreted onto the WD during this period would be burnt to He and simply be added to the mass of the WD.  Irrespective of the net gain or loss of accumulated mass during the nova eruption, `refuelling' would enable net WD mass growth over the refuelling period.  Here we only offer an outline of the concept, this is explored in more detail, observationally and theoretically, in \citet{12a2016}.

\subsection{The donor}\label{sec:donor}

There is an expectation that NIR observations of a quiescent nova system will largely isolate the donor star \cite[see][and Figure~\ref{fig:sed_q}]{2012ApJ...746...61D}; particularly for evolved (i.e.\ luminous) donors.  The accretion disk models employed in this work only extend to 7500\,\AA, but a simple extrapolation to longer wavelengths confirms that we can expect little, or no, accretion contribution to the quiescent flux in the NIR regime.  Therefore we conclude that the PHAT NIR quiescent photometry {\it should} simply be photometry of the \novak\ mass donor.  In Section~\ref{donor_excess}, we used the accretion disk modeling at quiescence to estimate the $I$-band (F814W) contribution from the donor.

\begin{figure*}
\includegraphics[width=\columnwidth]{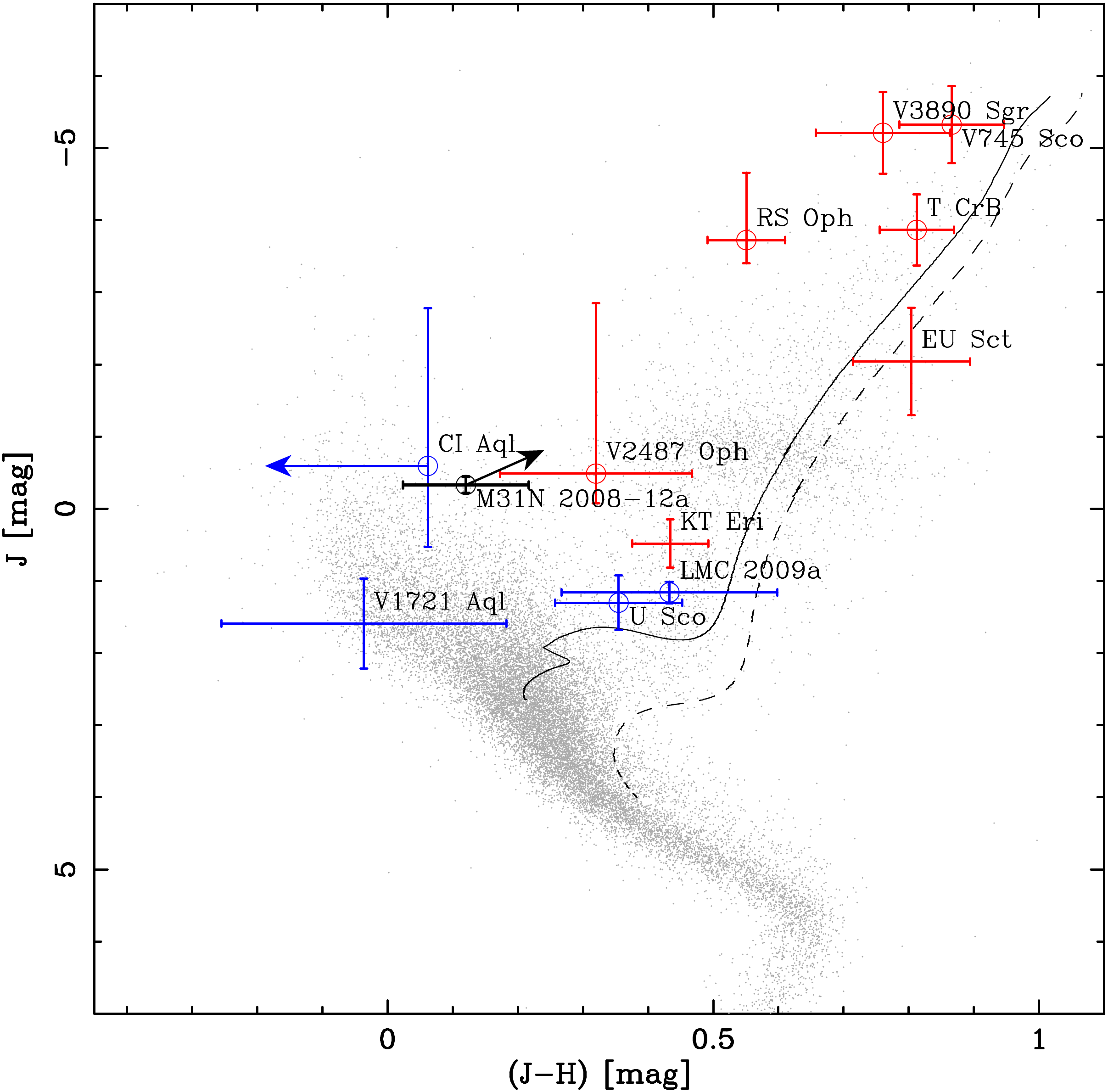}\hfill
\includegraphics[width=\columnwidth]{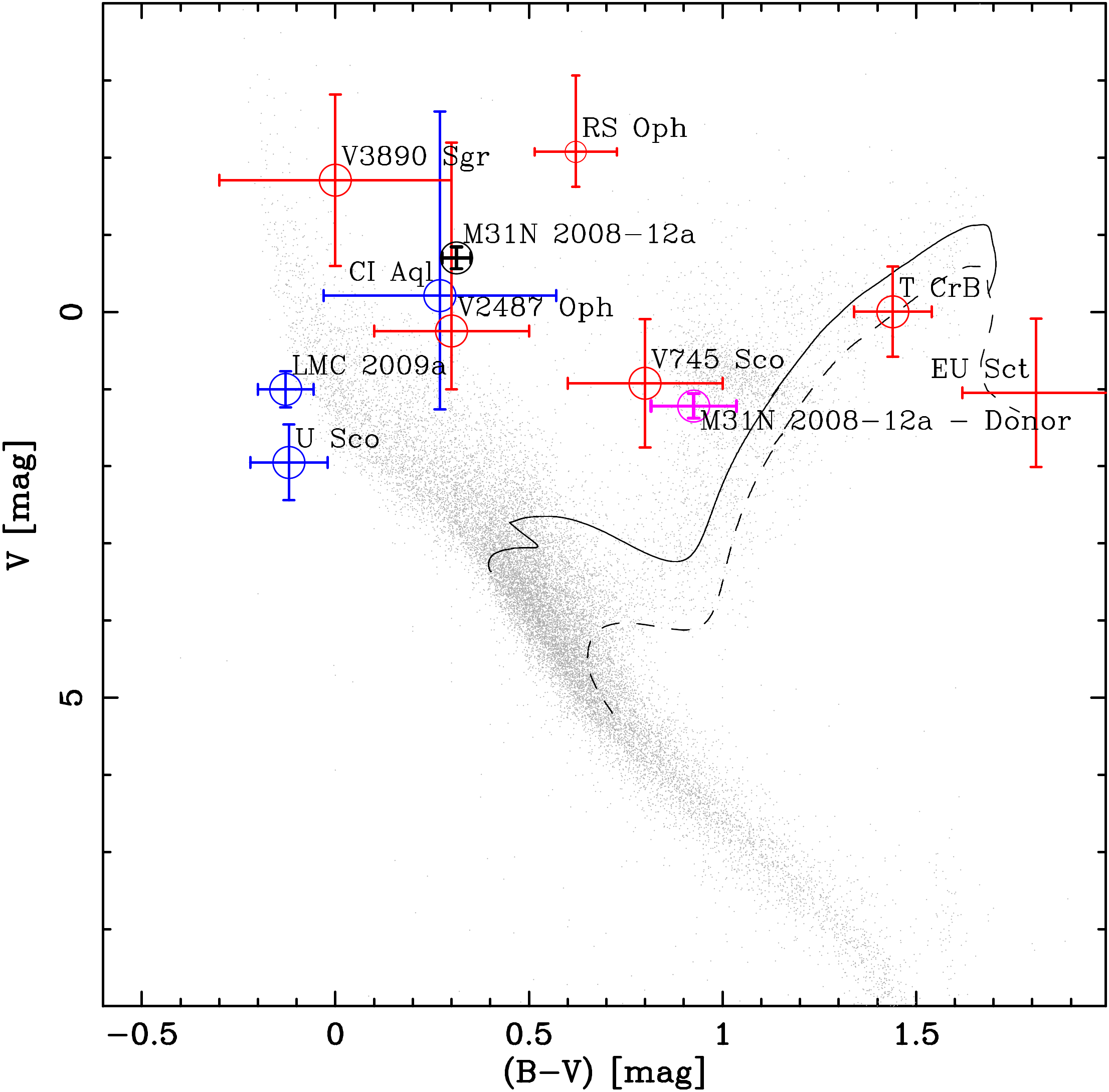}
\caption{Color--magnitude diagrams showing stars from the Hipparcos data set \citep[gray points;][]{1997A&A...323L..49P} with parallax errors $<10\%$. These stars have been transformed to the distance and extinction of \novak. Photometry is taken directly from Hipparcos or the the 2MASS catalog \citep{2006AJ....131.1163S}.  The hashed-black and solid-black lines are evolutionary tracks of $1\,M_\odot$ and $1.4\,M_\odot$ stars, respectively \citep{2004ApJ...612..168P}. The red points represent Galactic RG-novae and the blue points Galactic or LMC SG-novae \citep[see][and references therein]{2010ApJS..187..275S,2012ApJ...746...61D,2016ApJ...818..145B}. The known RNe in this sample have been identified by an additional circle.  {\bf Left:} NIR color--magnitude diagram, the black data point shows the assumed location of the \novak\ quiescent system at minimum ($t\sim75$\,d), considering the uncertainty in the photometry and extinction, the black arrow indicates the increase in emission during quiescence ($t\sim270$\,d). {\bf Right:} standard color-magnitude diagram showing the both position of the quiescent \novak\ system (black point; accretion disk+donor; $t\sim75$\,d) and the inferred position of the mass donor alone (magenta), which is consistent with the M\,31 red clump (see Section~\ref{sec:donor}).\label{fig:new}}
\end{figure*}

As can be seen in the left plot of Figure~\ref{fig:new}, the position of the quiescent \novak\ on a NIR color--magnitude diagram indicates that the donor is significantly less luminous and bluer than the red giants contained in the Galactic RG-novae (red points).  However, the \novak\ donor may be consistent with the M\,31 red clump.  A simple black body fit to the F110W ($\sim J$) and F160W ($\sim H$) photometry at quiescence ($t\sim75$\,d) yields $L_\mathrm{donor}\sim100\,L_\odot$ and $T_\mathrm{eff,donor}\sim4800$\,K.

If we include the extrapolated $I$-band luminosity of the donor (see Section~\ref{donor_excess}), then the three-point donor SED is very well represented by the same black body fit.  Hence we find that the \novak\ donor may be consistent with a black body of $T_\mathrm{eff, donor}=4890\pm110$\,K, $L_\mathrm{donor}=103^{+12}_{-11}\,L_\odot$, and $R_\mathrm{donor}=14.14^{+0.46}_{-0.47}\,R_\odot$, at the distance of M\,31.  This black body fit is illustrated by the dashed black line in the right-hand plot of Figure~\ref{fig:sed_q}.  We note that the quoted uncertainties are the formal $1\sigma$ errors resulting from the fitting process, the effect of possible systematic uncertainties related to the accretion and black-body disk modeling have not been estimated.  Using this black body, we compute the extrapolated $B$ and $V$ photometry (again indicating the above caveats) of the donor and plot it's position on a standard color--magnitude diagram in the right plot of Figure~\ref{fig:new}.  Here, it is clear that the donor may indeed be consistent with the M\,31 red clump.

As is illustrated in the left plot of Figure~\ref{fig:new}, as the quiescent system evolves from its minimum state ($t\sim75$\,days) toward the next eruption ($t\sim270$\,days) the donor brightens by $J\sim1$\,mag, but becomes redder -- consistent with an increase in the donor radius.  Such behavior may be related to irradiation of the donor, causing heating and expansion of the atmosphere.  It could also be related to the orbital phase of the system, with a tidally locked donor, for example, being non-spherically symmetric and unevenly heated -- but such phase effects would imply a high system inclination.  The very high luminosity of the accretion disk almost certainly now rules out high inclinations.

Given the available evidence, we must conclude that the mass donor in the \novak\ system is either a (low luminosity) red giant or post red giant branch star (e.g.\ horizontal branch), and/or that it is affected by significant irradiation from the primary, the disk, and the eruptions.  Of course we must point out that there is a possibility that the star identified as the donor may simply be another star at a very similar position on the sky within M\,31.  If we were relying on WFC/IR photometry of the donor alone, this probability would be quite large, but given the F814W spatial resolution of {\it HST} , the likelihood will be relatively small (around 2\%, see \oonek\ and \citealt{2014ApJS..213...10W}).

One piece of evidence may be key to constraining the donor however.  \otwok\ proposed that the \novak\ ejecta interacts strongly, and immediately, with material in the circumbinary medium with a $1/r^{2}$ density dependence.  This picture was strengthened by the reanalysis presented in \othreek.  With such behavior being seen consistently across four consecutive eruptions (2012--2015), the circumbinary material must be continuously replenished.  One feasible source of such material seems to be a stellar wind from a red giant donor as observed in RS\,Oph \citep[see e.g.,][]{2006ApJ...652..629B}.  As ejecta--circumbinary shocks are not observed in CNe, we must infer that Roche-lobe overflow is too efficient a mass transfer process to build up significant material in the circumbinary environment.  Therefore the donor would be strongly constrained to any star capable of generating such a wind.

However, in this paper we have explored the possibility that the extremely luminous accretion disk generates a significant disk wind.  Therefore, is it possible that such a disk wind is the source of the circumbinary pollution, not the stellar wind of a red giant donor?  As such a giant donor could be transferring matter at a high rate to the disk via Roche lobe overflow.  Indeed, could such a scenario be the only feasible manner in which such a high sustained WD mass accretion rate could be achieved?

\subsection{Orbital Period}\label{orb_per}

To date, the orbital period of the \novak\ system has eluded observation.  However, given we now know the mass of the WD and have constrained the radius of the donor, we can place some restrictions on $P_\mathrm{orb}$.  We will assume that the donor has evolved at least enough to reside on the red giant branch and that it was originally the lower mass component of the binary.  Therefore the donor mass must be somewhere in the range $0.8-8\,M_\odot$\footnote{The lower limit to allow evolution onto the red giant branch by the present day, the upper is the approximate zero-age upper mass limit to form a WD.}.  If we assume that the donor is Roche lobe filling, then the orbital separation must be in the range $25-44\,R_\odot$ ($0.12-0.20$\,AU), hence $5\lesssim P_\mathrm{orb}\lesssim23$\,days.  If the accretion is stellar wind driven then  $P_\mathrm{orb}\gg5$\,days.  We note that such minimum orbital separations, and hence Roche lobe sizes, are far too large to account for natural accretion disk truncation by the presence of the donor.

Based on a suspected red giant donor, \othreek\ suggested that \novak\ might be the only known nova with $P_\mathrm{rec}<P_\mathrm{orb}$.  Therefore, we again point out that as $\dot{M}$ is a function of the donor--WD separation, that if $P_\mathrm{rec}<P_\mathrm{orb}$ any orbital eccentricity may affect the accretion rate and inter eruption timescale on an eruption by eruption basis.  

\subsection{The X-ray flash non-detection}

The production of an X-ray flash at the onset of a nova eruption is a long standing prediction \citep{1990LNP...369..306S,2002AIPC..637..345K}. \citet{2016ApJ...830...40K} reported the results of an intensive {\it Swift} observing campaign to detect any X-ray flash associated with the 2015 eruption of \novak.  This campaign did not detect such a flash and \citet{2016ApJ...830...40K} presented two explanations for the non-detection.  Firstly, that the X-ray flash simply occurred before the {\it Swift} monitoring began -- which requires the X-ray flash to precede the optical/NUV nova by $\gtrsim8$\,days.  The second proposed that significant circumbinary material masked the flash signature.  At the time, with little firm evidence for the nature of the donor, \citet{2016ApJ...830...40K} preferred the first explanation.  However, given the work reported in this paper, we emphasize that the X-ray flash could have been missed due to significant absorption from circumbinary material.  This material could consist of some combination of a donor wind and a disk wind.  For a low inclination system with a significant disk wind, the bulk of the circumbinary material could even reside along the line of sight.  But the material in a disk wind dominated scenario may be expected to already be highly ionised, and hence unlikely to be able to mask any flash.  Therefore, the X-ray flash could have been absorbed if there was significant pollution of the cicumbinary environment by the wind of the donor, but likely only if the donor isn't Roche lobe filling.

\subsection{Time to reach the Chandrasekhar mass}

\ponek\ presented a prediction of the time required for the WD within the \novak\ system to grow to the Chandrasekhar mass (or at least to 1.37\,M$_\odot$).  For example, they presented a 1.36\,M$_\odot$ WD, with $\dot{M}_\mathrm{acc}=1.7\times10^{-7}\,M_\odot\,\mathrm{yr}^{-1}$, and a mass retention rate (the amount of accreted material remaining on the WD surface post-eruption) of 35\%.  This resulted in a time scale to grow to the Chandrasekhar mass of $\sim200$\,kyr.

We can update this simple calculation using the results from this paper and from \otwok, but using the same approach as \ponek.  We will assume that the WD mass is actually 1.37\,M$_\odot$, as used for the disk modeling and that a further 0.01\,M$_\odot$ of accretion is required to reach the Chandrasekhar mass (the same required mass growth as \ponek).  \xtwok\ determined that $\left(2.6\pm0.4\right)\times10^{-8}\,M_\odot$ of H is ejected in each eruption -- as a conservative estimate we will therefore assume that the total ejected mass is $6\times10^{-8}\,M_\odot$\footnote{Assuming roughly equal mass of H and He in the ejecta.  \citet{2014ApJ...793..136K} assume X=0.55, Y=0.43, and Z=0.02 in the ejecta.} \citep[consistent with][]{2015ApJ...808...52K}\footnote{We further note that this ejected mass assumes a spherical geometry.  With highly asymmetrical ejecta and the proposed low inclination, it is possible that the ejected mass is much greater.}.  Retaining this conservative stance, we will assume that the {\it average} WD mass accretion rate over the entire 1\,yr cycle is in fact the absolute {\it minimum} rate predicted by this paper, $\dot{M}_\mathrm{acc}=6.4\times10^{-7}\,M_\odot\,\mathrm{yr}^{-1}$; assuming a similar amount of mass is lost in the form of a disk wind.  Even then at such a lower accretion limit, the retained mass, or accretion efficiency is a staggeringly high 90\%.  Much higher than the $\sim30\%$ predicted by \ponek\ and the 63\% calculated by \citet{2015ApJ...808...52K}.  Combining these new data, we arrive at an updated prediction of the time to reach the Chandrasekhar mass of $<20$\,kyr ---  {\it possibly} much shorter.  

\section{Summary \& Conclusions}\label{sec:conc}

In this paper we have presented our analysis of an unrivalled series of {\it Hubble Space Telescope} photometric observations of the final stages of the 2015 eruption of \novak.  In this analysis, we have also exploited archival {\it HST} imaging during quiescence, and Keck spectroscopy 2014 eruption of \novak.  Our main findings include:

\begin{enumerate}
\item The {\it HST} WFC3/UVIS photometry of the late decline of the 2015 eruption shows a steady decline toward quiescence from the $\sim I$-band to the NUV, broadly consistent with the general trends established from the early-decline ground-based and {\it Swift} photometry.
\item When combined with the archival {\it HST} photometry -- shown to have been taken between eruptions -- the system appears to reach a flux minimum (in all bands) $\sim75$\,d post-eruption, before again increasing in luminosity by $\sim270$\,d post-eruption; $\sim100$\,d before the next eruption.
\item The broadband SEDs of the late decline and during quiescence were explored using accretion disk models.  The results indicate that these SEDs, even as early as 13.4\,d post-eruption, are consistent with the emission being dominated by an accretion disk -- one that has survived the eruption.
\item The inferred accretion rates are initially above the Eddington accretion limit ($t=13.4$\,d), indicating a disk that has survived albeit in a severely shocked and heated state.  The disk luminosity and inferred $\dot{M}$ then decline toward minimum ($t\sim75$\,d) before increasing again ($t\sim270$\,d) presumably as the disk fully reestablishes.  
\item The computed accretion rates, even at quiescence are large, with the disk luminosities still close to the Eddington limit. We speculate that mass loss from the disk will lead to a disk wind.
\item Could such a disk wind contribute a significant quantity of material to the circumbinary environment, and could provide the matter source with which the ejecta are observed to interact, and possibly even the ejecta collimation mechanism?
\item Our disk modelling computed a range of $\dot{M}=(1.2-2.8)\times10^{-6}\,M_\odot\,\mathrm{yr}^{-1}$ during quiescence.  Even when accounting for disk winds, which might account for half of $\dot{M}$, the derived accretion rates onto the WD at quiescence are still in the range $\dot{M}_\mathrm{acc}=(0.6-1.4)\times10^{-6}\,M_\odot\,\mathrm{yr}^{-1}$, significantly larger than any other nova.  If confirmed, WD accretion rates this high cause significant problems for a number of well-established nova eruption models.
\item \othreek\ and \hstk\ both proposed the presence of highly collimated outflows or even jets during the eruption.  Could such a high $\dot{M}$ drive a bipolar outflow from the inner disk and boundary layer, even at quiescence? 
\item Archival {\it HST} WFC3/IR photometry on the system isolates the donor.  Coupled with a strong $\sim I$-band excess from the accretion disk modeling, this photometry indicates a donor with $T_\mathrm{eff, donor}=4890\pm110$\,K, $L_\mathrm{donor}=103^{+12}_{-11}\,L_\odot$, and $R_\mathrm{donor}=14.14^{+0.46}_{-0.47}\,R_\odot$ -- consistent with the M\,31 red clump.
\item The NIR colors of the donor are slightly redward of the red clump and there is significant variation in the donor luminosity at quiescence.  These may be signs that the donor is significantly irradiated by the WD, disk, and ejecta, or may also be due to orbital phase affects.
\item Based on the work presented in this paper, the updated time-scale for the system to reach the Chandrasekhar mass has fallen to $<20$\,kyr.
\end{enumerate}

These {\it HST} observations of the late-decline of the 2015 eruption, combined with serendipitous archival detections during quiescence have started to shed some light on the inter-eruption behaviour of \novak.  It is clear that UV observations of this remarkable system are key to fully entangling the extreme physics at play throughout the entire eruption cycle.  Vital questions that should be addressed over the coming eruptions include the balance between accreted matter and ejected matter, in the light of the apparent large variation in quiescent $\dot{M}$ -- to fully assess the ultimate fate of \novak.

\begin{acknowledgements}

We would firstly like to express our gratitude to the anonymous referee for their detailed comments and suggestions, which helped improve the content and readability of the manuscript.

Based on observations made with the NASA/ESA Hubble Space Telescope, obtained from the Data Archive at the Space Telescope Science Institute, which is operated by the Association of Universities for Research in Astronomy, Inc., under NASA contract NAS 5-26555. These observations are associated with programs \#12056, \#12106, and \#14125.

The W.\ M.\ Keck Observatory is operated as a scientific partnership among the California Institute of Technology, the University of California and the National Aeronautics and Space Administration. The Observatory was made possible by the generous financial support of the W. M. Keck Foundation.  The authors wish to recognize the significant cultural role that the summit of Maunakea has always had within the indigenous Hawai'ian community.  We are most fortunate to have the opportunity to conduct observations from this mountain.

This research has made use of the Keck Observatory Archive (KOA), which is operated by the W.\,M.\ Keck Observatory and the NASA Exoplanet Science Institute (NExScI), under contract with the National Aeronautics and Space Administration.  The authors also acknowledge Sumin Tang as the PI of program C204LA.  We also acknowledge the Keck observer Adam Miller.

The authors would like to thank the {\it HST} staff for their heroic efforts scheduling our early-time, and disruptive, spectroscopic observations.  Particular thanks are given to Charles R.\ Proffitt, the STScI Contact Scientist for programme \#14125, for his support with the STIS observations.

MJD would like to personally thank Mariko Kato for providing Figure~\ref{kato_plot}, and Christian Knigge and James Matthews for discussions about accretion disk winds.

PG wishes to thank William (Bill) P.\ Blair for his kind hospitality in the Rowland Department of Physics \& Astronomy at the Johns Hopkins University.

MH acknowledges the support of the Spanish Ministry of Economy and Competitiveness (MINECO) under the grant FDPI-2013-16933 as well as the support of the Generalitat de Catalunya/CERCA programme. 

SCW acknowledges a visiting research fellowship at Liverpool John Moores University. 

KH was supported by the project RVO:67985815.

VARMR acknowledges partial financial support from the Radboud Excellence Initiative, from Funda\c{c}\~ao para a Ci\^encia e a Tecnologia (FCT) in the form of an exploratory project of reference IF/00498/2015, from Center for Research \& Development in Mathematics and Applications (CIDMA) strategic project UID/MAT/04106/2013 and from Enabling Green E-science for the Square Kilometer Array Research Infrastructure (ENGAGE SKA), POCI-01-0145-FEDER-022217, funded by Programa Operacional Competitividade e Internacionaliza\c{c}\~{a}o (COMPETE 2020) and FCT, Portugal.

This work has been supported in part by NSF grant AST-1009566.  Support for program \#14125 was provided by NASA through a grant from the Space Telescope Science Institute, which is operated by the Association of Universities for Research in Astronomy, Inc., under NASA contract NAS 5-26555.
\end{acknowledgements}

\facilities{HST (WFC3), Keck:I (LRIS)}

\software{calwf3 pipeline \citep[v3.1.6;][]{2012wfci.book.....D}, Dolphot \citep[v2.0;][]{2000PASP..112.1383D}, Drizzlepac (v2.0.2), IRAF \citep[v2.16.1,][]{1993ASPC...52..173T}, PGPLOT (v5.2), Synspec / Tlusty \citep[v202;][]{1995ApJ...439..875H}, {\tt wfc3uv\_ctereverse\_parallel} \citep[v2015.07.22;][]{CTE}}
	
\bibliographystyle{aasjournal} 
\bibliography{refs} 

\begin{thebibliography}{}
\expandafter\ifx\csname natexlab\endcsname\relax\def\natexlab#1{#1}\fi

\bibitem[{{Abramowicz} {et~al.}(1988){Abramowicz}, {Czerny}, {Lasota}, \&
  {Szuszkiewicz}}]{1988ApJ...332..646A}
{Abramowicz}, M.~A., {Czerny}, B., {Lasota}, J.~P., \& {Szuszkiewicz}, E. 1988,
  \apj, 332, 646

\bibitem[{{Anderson} {et~al.}(2012){Anderson}, {MacKenty}, {Baggett}, \&
  {Noeske}}]{CTE}
{Anderson}, J., {MacKenty}, J., {Baggett}, S., \& {Noeske}, K. 2012, {The
  Efficacy of Post-Flashing for Mitigating CTE-Losses in WFC3/UVIS Images},
  {STScI},
  \url{http://www.stsci.edu/hst/wfc3/ins_performance/CTE/ANDERSON_UVIS_POSTFLASH_EFFICACY.pdf}

\bibitem[{{Ashall} {et~al.}(2016){Ashall}, {Mazzali}, {Sasdelli}, \&
  {Prentice}}]{2016MNRAS.460.3529A}
{Ashall}, C., {Mazzali}, P., {Sasdelli}, M., \& {Prentice}, S.~J. 2016, \mnras,
  460, 3529

\bibitem[{{Barsukova} {et~al.}(2011){Barsukova}, {Fabrika}, {Hornoch},
  {Fatkhullin}, {Sholukhova}, \& {Pietsch}}]{2011ATel.3725....1B}
{Barsukova}, E., {Fabrika}, S., {Hornoch}, K., {et~al.} 2011, The Astronomer's
  Telegram, 3725, 1

\bibitem[{{Bode} \& {Evans}(2008)}]{2008clno.book.....B}
{Bode}, M.~F., \& {Evans}, A., eds. 2008, Cambridge Astrophysics Series,
  Vol.~43, {Classical Novae, 2nd Edition} (Cambridge: Cambridge University
  Press)

\bibitem[{{Bode} \& {Kahn}(1985)}]{1985MNRAS.217..205B}
{Bode}, M.~F., \& {Kahn}, F.~D. 1985, \mnras, 217, 205

\bibitem[{{Bode} {et~al.}(2006){Bode}, {O'Brien}, {Osborne}, {Page},
  {Senziani}, {Skinner}, {Starrfield}, {Ness}, {Drake}, {Schwarz}, {Beardmore},
  {Darnley}, {Eyres}, {Evans}, {Gehrels}, {Goad}, {Jean}, {Krautter}, \&
  {Novara}}]{2006ApJ...652..629B}
{Bode}, M.~F., {O'Brien}, T.~J., {Osborne}, J.~P., {et~al.} 2006, \apj, 652,
  629

\bibitem[{{Bode} {et~al.}(2016){Bode}, {Darnley}, {Beardmore}, {Osborne},
  {Page}, {Walter}, {Krautter}, {Melandri}, {Ness}, {O'Brien}, {Orio},
  {Schwarz}, {Shara}, \& {Starrfield}}]{2016ApJ...818..145B}
{Bode}, M.~F., {Darnley}, M.~J., {Beardmore}, A.~P., {et~al.} 2016, \apj, 818,
  145

\bibitem[{{Cao} {et~al.}(2012){Cao}, {Kasliwal}, {Neill}, {Kulkarni}, {Lou},
  {Ben-Ami}, {Bloom}, {Cenko}, {Law}, {Nugent}, {Ofek}, {Poznanski}, \&
  {Quimby}}]{2012ApJ...752..133C}
{Cao}, Y., {Kasliwal}, M.~M., {Neill}, J.~D., {et~al.} 2012, \apj, 752, 133

\bibitem[{{Dalcanton} {et~al.}(2012){Dalcanton}, {Williams}, {Lang}, {Lauer},
  {Kalirai}, {Seth}, {Dolphin}, {Rosenfield}, {Weisz}, {Bell}, {Bianchi},
  {Boyer}, {Caldwell}, {Dong}, {Dorman}, {Gilbert}, {Girardi}, {Gogarten},
  {Gordon}, {Guhathakurta}, {Hodge}, {Holtzman}, {Johnson}, {Larsen}, {Lewis},
  {Melbourne}, {Olsen}, {Rix}, {Rosema}, {Saha}, {Sarajedini}, {Skillman}, \&
  {Stanek}}]{2012ApJS..200...18D}
{Dalcanton}, J.~J., {Williams}, B.~F., {Lang}, D., {et~al.} 2012, \apjs, 200,
  18

\bibitem[{{Darnley} {et~al.}(2015{\natexlab{a}}){Darnley}, {Henze}, {Shafter},
  \& {Kato}}]{2015ATel.7964....1D}
{Darnley}, M.~J., {Henze}, M., {Shafter}, A.~W., \& {Kato}, M.
  2015{\natexlab{a}}, The Astronomer's Telegram, 7964, 1

\bibitem[{{Darnley} {et~al.}(2015{\natexlab{b}}){Darnley}, {Henze}, {Shafter},
  \& {Kato}}]{2015ATel.7965....1D}
---. 2015{\natexlab{b}}, The Astronomer's Telegram, 7965, 1

\bibitem[{{Darnley} {et~al.}(2008){Darnley}, {Hounsell}, \&
  {Bode}}]{2008ASPC..401..203D}
{Darnley}, M.~J., {Hounsell}, R.~A., \& {Bode}, M.~F. 2008, in Astronomical
  Society of the Pacific Conference Series, Vol. 401, RS Ophiuchi (2006) and
  the Recurrent Nova Phenomenon, ed. A.~{Evans}, M.~F. {Bode}, T.~J. {O'Brien},
  \& M.~J. {Darnley}, 203

\bibitem[{{Darnley} {et~al.}(2012){Darnley}, {Ribeiro}, {Bode}, {Hounsell}, \&
  {Williams}}]{2012ApJ...746...61D}
{Darnley}, M.~J., {Ribeiro}, V.~A.~R.~M., {Bode}, M.~F., {Hounsell}, R.~A., \&
  {Williams}, R.~P. 2012, \apj, 746, 61

\bibitem[{{Darnley} {et~al.}(2014){Darnley}, {Williams}, {Bode}, {Henze},
  {Ness}, {Shafter}, {Hornoch}, \& {Votruba}}]{2014A&A...563L...9D}
{Darnley}, M.~J., {Williams}, S.~C., {Bode}, M.~F., {et~al.} 2014, \aap, 563,
  L9

\bibitem[{{Darnley} {et~al.}(2015{\natexlab{c}}){Darnley}, {Henze}, {Steele},
  {Bode}, {Ribeiro}, {Rodr{\'{\i}}guez-Gil}, {Shafter}, {Williams}, {Baer},
  {Hachisu}, {Hernanz}, {Hornoch}, {Hounsell}, {Kato}, {Kiyota}, {Ku{\v
  c}{\'a}kov{\'a}}, {Maehara}, {Ness}, {Piascik}, {Sala}, {Skillen}, {Smith},
  \& {Wolf}}]{2015A&A...580A..45D}
{Darnley}, M.~J., {Henze}, M., {Steele}, I.~A., {et~al.} 2015{\natexlab{c}},
  \aap, 580, A45

\bibitem[{{Darnley} {et~al.}(2016){Darnley}, {Henze}, {Bode}, {Hachisu},
  {Hernanz}, {Hornoch}, {Hounsell}, {Kato}, {Ness}, {Osborne}, {Page},
  {Ribeiro}, {Rodr{\'{\i}}guez-Gil}, {Shafter}, {Shara}, {Steele}, {Williams},
  {Arai}, {Arcavi}, {Barsukova}, {Boumis}, {Chen}, {Fabrika}, {Figueira},
  {Gao}, {Gehrels}, {Godon}, {Goranskij}, {Harman}, {Hartmann}, {Hosseinzadeh},
  {Horst}, {Itagaki}, {Jos{\'e}}, {Kabashima}, {Kaur}, {Kawai}, {Kennea},
  {Kiyota}, {Ku{\v c}{\'a}kov{\'a}}, {Lau}, {Maehara}, {Naito}, {Nakajima},
  {Nishiyama}, {O'Brien}, {Quimby}, {Sala}, {Sano}, {Sion}, {Valeev},
  {Watanabe}, {Watanabe}, {Williams}, \& {Xu}}]{2016ApJ...833..149D}
{Darnley}, M.~J., {Henze}, M., {Bode}, M.~F., {et~al.} 2016, \apj, 833, 149

\bibitem[{Darnley {et~al.}(2017)Darnley, Hounsell, Godon, Perley, Henze, Kuin,
  Williams, Williams, Bode, Harman, Hornoch, Link, Ness, Ribeiro, Sion,
  Shafter, \& Shara}]{0004-637X-847-1-35}
Darnley, M.~J., Hounsell, R., Godon, P., {et~al.} 2017, The Astrophysical
  Journal, 847, 35

\bibitem[{{Dolphin}(2000)}]{2000PASP..112.1383D}
{Dolphin}, A.~E. 2000, \pasp, 112, 1383

\bibitem[{{Dressel}(2012)}]{2012wfci.book.....D}
{Dressel}, L. 2012, {Wide Field Camera 3 Instrument Handbook for Cycle 21 v.
  5.0} (Baltimore, MD: STScI)

\bibitem[{{Ellis} {et~al.}(1984){Ellis}, {Grayson}, \&
  {Bond}}]{1984PASP...96..283E}
{Ellis}, G.~L., {Grayson}, E.~T., \& {Bond}, H.~E. 1984, \pasp, 96, 283

\bibitem[{{Evans} {et~al.}(2008){Evans}, {Bode}, {O'Brien}, \&
  {Darnley}}]{2008ASPC..401.....E}
{Evans}, A., {Bode}, M.~F., {O'Brien}, T.~J., \& {Darnley}, M.~J., eds. 2008,
  Astronomical Society of the Pacific Conference Series, Vol. 401, {RS Ophiuchi
  (2006) and the Recurrent Nova Phenomenon} (San Francisco: Astronomical
  Society of the Pacific)

\bibitem[{{Fujimoto}(1982)}]{1982ApJ...257..752F}
{Fujimoto}, M.~Y. 1982, \apj, 257, 752

\bibitem[{{Godon}(1997)}]{1997ApJ...483..882G}
{Godon}, P. 1997, \apj, 483, 882

\bibitem[{{Godon} {et~al.}(2017){Godon}, {Sion}, {Balman}, \&
  {Blair}}]{2017ApJ...846...52G}
{Godon}, P., {Sion}, E.~M., {Balman}, {\c S}., \& {Blair}, W.~P. 2017, \apj,
  846, 52

\bibitem[{{Godon} {et~al.}(2014){Godon}, {Sion}, {Starrfield}, {Livio},
  {Williams}, {Woodward}, {Kuin}, \& {Page}}]{2014ApJ...784L..33G}
{Godon}, P., {Sion}, E.~M., {Starrfield}, S., {et~al.} 2014, \apjl, 784, L33

\bibitem[{{Hachisu} \& {Kato}(2006)}]{2006ApJS..167...59H}
{Hachisu}, I., \& {Kato}, M. 2006, \apjs, 167, 59

\bibitem[{{Hachisu} \& {Kato}(2007)}]{2007ApJ...662..552H}
---. 2007, \apj, 662, 552

\bibitem[{{Hachisu} {et~al.}(2016){Hachisu}, {Saio}, \&
  {Kato}}]{2016ApJ...824...22H}
{Hachisu}, I., {Saio}, H., \& {Kato}, M. 2016, \apj, 824, 22

\bibitem[{{Hachisu} {et~al.}(2008){Hachisu}, {Kato}, {Kiyota}, {Kubotera},
  {Maehara}, {Nakajima}, {Ishii}, {Kamada}, {Mizoguchi}, {Nishiyama},
  {Sumitomo}, {Tanaka}, {Yamanaka}, \& {Sadakane}}]{2008ASPC..401..206H}
{Hachisu}, I., {Kato}, M., {Kiyota}, S., {et~al.} 2008, in Astronomical Society
  of the Pacific Conference Series, Vol. 401, RS Ophiuchi (2006) and the
  Recurrent Nova Phenomenon, ed. A.~{Evans}, M.~F. {Bode}, T.~J. {O'Brien}, \&
  M.~J. {Darnley}, 206

\bibitem[{{Henze} {et~al.}(2015{\natexlab{a}}){Henze}, {Darnley}, {Kabashima},
  {Nishiyama}, {Itagaki}, \& {Gao}}]{2015A&A...582L...8H}
{Henze}, M., {Darnley}, M.~J., {Kabashima}, F., {et~al.} 2015{\natexlab{a}},
  \aap, 582, L8

\bibitem[{{Henze} {et~al.}(2014){Henze}, {Ness}, {Darnley}, {Bode}, {Williams},
  {Shafter}, {Kato}, \& {Hachisu}}]{2014A&A...563L...8H}
{Henze}, M., {Ness}, J.-U., {Darnley}, M.~J., {et~al.} 2014, \aap, 563, L8

\bibitem[{{Henze} {et~al.}(2015{\natexlab{b}}){Henze}, {Ness}, {Darnley},
  {Bode}, {Williams}, {Shafter}, {Sala}, {Kato}, {Hachisu}, \&
  {Hernanz}}]{2015A&A...580A..46H}
---. 2015{\natexlab{b}}, \aap, 580, A46

\bibitem[{{Henze} {et~al.}(2015{\natexlab{c}}){Henze}, {Darnley}, {Shafter},
  {Kato}, {Hachisu}, {Bode}, {Ness}, {Osborne}, {Kennea}, \&
  {Gehrels}}]{2015ATel.7984....1H}
{Henze}, M., {Darnley}, M.~J., {Shafter}, A.~W., {et~al.} 2015{\natexlab{c}},
  The Astronomer's Telegram, 7984, 1

\bibitem[{{Henze} {et~al.}(2017{\natexlab{a}}){Henze}, {Darnley}, {Williams},
  {Hounsell}, {Bode}, {Harman}, {Hornoch}, {Ness}, {Ribeiro}, {Shafter}, \&
  {Shara}}]{HenzeQ}
{Henze}, M., {Darnley}, M.~J., {Williams}, S.~C., {et~al.} 2017{\natexlab{a}},
  in preparation, for submission to A\&A

\bibitem[{{Henze} {et~al.}(2017{\natexlab{b}}){Henze}, {Darnley}, {Williams},
  {Hounsell}, {Bode}, {Harman}, {Hornoch}, {Ness}, {Ribeiro}, {Shafter}, \&
  {Shara}}]{12a2016}
---. 2017{\natexlab{b}}, in preparation, for submission to ApJ

\bibitem[{{Hillman} {et~al.}(2016){Hillman}, {Prialnik}, {Kovetz}, \&
  {Shara}}]{2016ApJ...819..168H}
{Hillman}, Y., {Prialnik}, D., {Kovetz}, A., \& {Shara}, M.~M. 2016, \apj, 819,
  168

\bibitem[{{Hillwig} {et~al.}(2004){Hillwig}, {Livio}, \&
  {Honeycutt}}]{2004PASP..116..397H}
{Hillwig}, T., {Livio}, M., \& {Honeycutt}, R.~K. 2004, \pasp, 116, 397

\bibitem[{{Hubeny}(1988)}]{1988CoPhC..52..103H}
{Hubeny}, I. 1988, Computer Physics Communications, 52, 103

\bibitem[{{Hubeny} \& {Lanz}(1995)}]{1995ApJ...439..875H}
{Hubeny}, I., \& {Lanz}, T. 1995, \apj, 439, 875

\bibitem[{{Hubeny} {et~al.}(1994){Hubeny}, {Lanz}, \& {Jeffrey}}]{hub94}
{Hubeny}, I., {Lanz}, T., \& {Jeffrey}, S. 1994, St. Andrews Univ. Newsletter
  on Analysis of Astronomical Spectra, 20, 30

\bibitem[{{Itagaki} {et~al.}(2016){Itagaki}, {Gao}, {Darnley}, {Henze},
  {Shafter}, {Williams}, {Kafka}, {Kato}, \& {et al.}}]{2016ATel.9848....1I}
{Itagaki}, K., {Gao}, X., {Darnley}, M.~J., {et~al.} 2016, The Astronomer's
  Telegram, 9848

\bibitem[{{Kato} {et~al.}(2015){Kato}, {Saio}, \&
  {Hachisu}}]{2015ApJ...808...52K}
{Kato}, M., {Saio}, H., \& {Hachisu}, I. 2015, \apj, 808, 52

\bibitem[{{Kato} {et~al.}(2017{\natexlab{a}}){Kato}, {Saio}, \&
  {Hachisu}}]{2017ApJ...844..143K}
---. 2017{\natexlab{a}}, \apj, 844, 143

\bibitem[{{Kato} {et~al.}(2017{\natexlab{b}}){Kato}, {Saio}, \&
  {Hachisu}}]{2017ApJ...838..153K}
---. 2017{\natexlab{b}}, \apj, 838, 153

\bibitem[{{Kato} {et~al.}(2014){Kato}, {Saio}, {Hachisu}, \&
  {Nomoto}}]{2014ApJ...793..136K}
{Kato}, M., {Saio}, H., {Hachisu}, I., \& {Nomoto}, K. 2014, \apj, 793, 136

\bibitem[{{Kato} {et~al.}(2016){Kato}, {Saio}, {Henze}, {Ness}, {Osborne},
  {Page}, {Darnley}, {Bode}, {Shafter}, {Hernanz}, {Gehrels}, {Kennea}, \&
  {Hachisu}}]{2016ApJ...830...40K}
{Kato}, M., {Saio}, H., {Henze}, M., {et~al.} 2016, \apj, 830, 40

\bibitem[{{King}(1998)}]{1998MNRAS.296L..45K}
{King}, A.~R. 1998, \mnras, 296, L45

\bibitem[{{Knigge} {et~al.}(2000){Knigge}, {King}, \&
  {Patterson}}]{2000A&A...364L..75K}
{Knigge}, C., {King}, A.~R., \& {Patterson}, J. 2000, \aap, 364, L75

\bibitem[{{Korotkiy} \& {Elenin}(2011)}]{2011Kor}
{Korotkiy}, S., \& {Elenin}, L. 2011, {CBAT},  {IAU},
  \url{http://www.cbat.eps.harvard.edu/unconf/followups/J00452885+4154094.html}

\bibitem[{{Krautter}(2002)}]{2002AIPC..637..345K}
{Krautter}, J. 2002, in American Institute of Physics Conference Series, Vol.
  637, Classical Nova Explosions, ed. M.~{Hernanz} \& J.~{Jos{\'e}}, 345--354

\bibitem[{{Livio}(1997)}]{1997ASPC..121..845L}
{Livio}, M. 1997, in Astronomical Society of the Pacific Conference Series,
  Vol. 121, IAU Colloq. 163: Accretion Phenomena and Related Outflows, ed.
  D.~T. {Wickramasinghe}, G.~V. {Bicknell}, \& L.~{Ferrario}, 845

\bibitem[{{Matthews} {et~al.}(2015){Matthews}, {Knigge}, {Long}, {Sim}, \&
  {Higginbottom}}]{2015MNRAS.450.3331M}
{Matthews}, J.~H., {Knigge}, C., {Long}, K.~S., {Sim}, S.~A., \&
  {Higginbottom}, N. 2015, \mnras, 450, 3331

\bibitem[{{McCarthy} {et~al.}(1998){McCarthy}, {Cohen}, {Butcher}, {Cromer},
  {Croner}, {Douglas}, {Goeden}, {Grewal}, {Lu}, {Petrie}, {Weng}, {Weber},
  {Koch}, \& {Rodgers}}]{1998SPIE.3355...81M}
{McCarthy}, J.~K., {Cohen}, J.~G., {Butcher}, B., {et~al.} 1998, in \procspie,
  Vol. 3355, Optical Astronomical Instrumentation, ed. S.~{D'Odorico}, 81--92

\bibitem[{{Narayan} \& {Yi}(1994)}]{1994ApJ...428L..13N}
{Narayan}, R., \& {Yi}, I. 1994, \apjl, 428, L13

\bibitem[{{Narayan} \& {Yi}(1995)}]{1995ApJ...444..231N}
---. 1995, \apj, 444, 231

\bibitem[{{Nishiyama} \& {Kabashima}(2008)}]{2008Nis}
{Nishiyama}, K., \& {Kabashima}, F. 2008, {CBAT},  {IAU},
  \url{http://www.cbat.eps.harvard.edu/iau/CBAT\_M31.html\#2008-12a}

\bibitem[{{Nishiyama} \& {Kabashima}(2012)}]{2012Nis}
---. 2012, {CBAT},  {IAU},
  \url{http://www.cbat.eps.harvard.edu/unconf/followups/J00452884+4154095.html}

\bibitem[{{Nomoto}(1982)}]{1982ApJ...253..798N}
{Nomoto}, K. 1982, \apj, 253, 798

\bibitem[{{Oke} {et~al.}(1995){Oke}, {Cohen}, {Carr}, {Cromer}, {Dingizian},
  {Harris}, {Labrecque}, {Lucinio}, {Schaal}, {Epps}, \&
  {Miller}}]{1995PASP..107..375O}
{Oke}, J.~B., {Cohen}, J.~G., {Carr}, M., {et~al.} 1995, \pasp, 107, 375

\bibitem[{{Page} {et~al.}(2015){Page}, {Osborne}, {Kuin}, {Henze}, {Walter},
  {Beardmore}, {Bode}, {Darnley}, {Delgado}, {Drake}, {Hernanz}, {Mukai},
  {Nelson}, {Ness}, {Schwarz}, {Shore}, {Starrfield}, \&
  {Woodward}}]{2015MNRAS.454.3108P}
{Page}, K.~L., {Osborne}, J.~P., {Kuin}, N.~P.~M., {et~al.} 2015, \mnras, 454,
  3108

\bibitem[{{Pagnotta} {et~al.}(2009){Pagnotta}, {Schaefer}, {Xiao}, {Collazzi},
  \& {Kroll}}]{2009AJ....138.1230P}
{Pagnotta}, A., {Schaefer}, B.~E., {Xiao}, L., {Collazzi}, A.~C., \& {Kroll},
  P. 2009, \aj, 138, 1230

\bibitem[{{Perryman} {et~al.}(1997){Perryman}, {Lindegren}, {Kovalevsky},
  {Hoeg}, {Bastian}, {Bernacca}, {Cr{\'e}z{\'e}}, {Donati}, {Grenon},
  {Grewing}, {van Leeuwen}, {van der Marel}, {Mignard}, {Murray}, {Le Poole},
  {Schrijver}, {Turon}, {Arenou}, {Froeschl{\'e}}, \&
  {Petersen}}]{1997A&A...323L..49P}
{Perryman}, M.~A.~C., {Lindegren}, L., {Kovalevsky}, J., {et~al.} 1997, \aap,
  323, L49

\bibitem[{{Pietrinferni} {et~al.}(2004){Pietrinferni}, {Cassisi}, {Salaris}, \&
  {Castelli}}]{2004ApJ...612..168P}
{Pietrinferni}, A., {Cassisi}, S., {Salaris}, M., \& {Castelli}, F. 2004, \apj,
  612, 168

\bibitem[{{Popham}(1997)}]{1997ApJ...478..734P}
{Popham}, R. 1997, \apj, 478, 734

\bibitem[{{Popham} \& {Narayan}(1995)}]{1995ApJ...442..337P}
{Popham}, R., \& {Narayan}, R. 1995, \apj, 442, 337

\bibitem[{{Poutanen} {et~al.}(2007){Poutanen}, {Lipunova}, {Fabrika},
  {Butkevich}, \& {Abolmasov}}]{2007MNRAS.377.1187P}
{Poutanen}, J., {Lipunova}, G., {Fabrika}, S., {Butkevich}, A.~G., \&
  {Abolmasov}, P. 2007, \mnras, 377, 1187

\bibitem[{{Pringle}(1977)}]{1977MNRAS.178..195P}
{Pringle}, J.~E. 1977, \mnras, 178, 195

\bibitem[{{Puebla} {et~al.}(2007){Puebla}, {Diaz}, \&
  {Hubeny}}]{2007AJ....134.1923P}
{Puebla}, R.~E., {Diaz}, M.~P., \& {Hubeny}, I. 2007, \aj, 134, 1923

\bibitem[{{Rockosi} {et~al.}(2010){Rockosi}, {Stover}, {Kibrick}, {Lockwood},
  {Peck}, {Cowley}, {Bolte}, {Adkins}, {Alcott}, {Allen}, {Brown}, {Cabak},
  {Deich}, {Hilyard}, {Kassis}, {Lanclos}, {Lewis}, {Pfister}, {Phillips},
  {Robinson}, {Saylor}, {Thompson}, {Ward}, {Wei}, \&
  {Wright}}]{2010SPIE.7735E..0RR}
{Rockosi}, C., {Stover}, R., {Kibrick}, R., {et~al.} 2010, in \procspie, Vol.
  7735, Ground-based and Airborne Instrumentation for Astronomy III, 77350R

\bibitem[{{Schaefer}(2010)}]{2010ApJS..187..275S}
{Schaefer}, B.~E. 2010, \apjs, 187, 275

\bibitem[{{Shafter}(2017)}]{2017ApJ...834..196S}
{Shafter}, A.~W. 2017, \apj, 834, 196

\bibitem[{{Shafter} {et~al.}(2012){Shafter}, {Hornoch}, {Ciardullo}, {Darnley},
  \& {Bode}}]{2012ATel.4503....1S}
{Shafter}, A.~W., {Hornoch}, K., {Ciardullo}, J.~V.~R., {Darnley}, M.~J., \&
  {Bode}, M.~F. 2012, The Astronomer's Telegram, 4503, 1

\bibitem[{{Shahbaz} \& {Kuulkers}(1998)}]{1998MNRAS.295L...1S}
{Shahbaz}, T., \& {Kuulkers}, E. 1998, \mnras, 295, L1

\bibitem[{{Shakura} \& {Sunyaev}(1973)}]{1973A&A....24..337S}
{Shakura}, N.~I., \& {Sunyaev}, R.~A. 1973, \aap, 24, 337

\bibitem[{{Skrutskie} {et~al.}(2006){Skrutskie}, {Cutri}, {Stiening},
  {Weinberg}, {Schneider}, {Carpenter}, {Beichman}, {Capps}, {Chester},
  {Elias}, {Huchra}, {Liebert}, {Lonsdale}, {Monet}, {Price}, {Seitzer},
  {Jarrett}, {Kirkpatrick}, {Gizis}, {Howard}, {Evans}, {Fowler}, {Fullmer},
  {Hurt}, {Light}, {Kopan}, {Marsh}, {McCallon}, {Tam}, {Van Dyk}, \&
  {Wheelock}}]{2006AJ....131.1163S}
{Skrutskie}, M.~F., {Cutri}, R.~M., {Stiening}, R., {et~al.} 2006, \aj, 131,
  1163

\bibitem[{Starrfield(2016)}]{Starrfield2016}
Starrfield, S. 2016, in Handbook of Supernovae, ed. A.~W. Alsabti \& P.~Murdin
  (Cham: Springer International Publishing), 1--26

\bibitem[{{Starrfield} {et~al.}(2016){Starrfield}, {Iliadis}, \&
  {Hix}}]{2016PASP..128e1001S}
{Starrfield}, S., {Iliadis}, C., \& {Hix}, W.~R. 2016, \pasp, 128, 051001

\bibitem[{{Starrfield} {et~al.}(1976){Starrfield}, {Sparks}, \&
  {Truran}}]{1976IAUS...73..155S}
{Starrfield}, S., {Sparks}, W.~M., \& {Truran}, J.~W. 1976, in IAU Symposium,
  Vol.~73, Structure and Evolution of Close Binary Systems, ed. P.~{Eggleton},
  S.~{Mitton}, \& J.~{Whelan}, 155--172

\bibitem[{{Starrfield} {et~al.}(1990){Starrfield}, {Truran}, {Sparks},
  {Krautter}, \& {MacDonald}}]{1990LNP...369..306S}
{Starrfield}, S., {Truran}, J.~W., {Sparks}, W.~M., {Krautter}, J., \&
  {MacDonald}, J. 1990, in Lecture Notes in Physics, Berlin Springer Verlag,
  Vol. 369, IAU Colloq. 122: Physics of Classical Novae, ed. A.~{Cassatella} \&
  R.~{Viotti}, 306

\bibitem[{{Tang} {et~al.}(2013){Tang}, {Cao}, \&
  {Kasliwal}}]{2013ATel.5607....1T}
{Tang}, S., {Cao}, Y., \& {Kasliwal}, M.~M. 2013, The Astronomer's Telegram,
  5607, 1

\bibitem[{{Tang} {et~al.}(2014){Tang}, {Bildsten}, {Wolf}, {Li}, {Kong}, {Cao},
  {Cenko}, {De Cia}, {Kasliwal}, {Kulkarni}, {Laher}, {Masci}, {Nugent},
  {Perley}, {Prince}, \& {Surace}}]{2014ApJ...786...61T}
{Tang}, S., {Bildsten}, L., {Wolf}, W.~M., {et~al.} 2014, \apj, 786, 61

\bibitem[{{Tody}(1993)}]{1993ASPC...52..173T}
{Tody}, D. 1993, in Astronomical Society of the Pacific Conference Series,
  Vol.~52, Astronomical Data Analysis Software and Systems II, ed. R.~J.
  {Hanisch}, R.~J.~V. {Brissenden}, \& J.~{Barnes}, 173

\bibitem[{{van Paradijs} \& {McClintock}(1994)}]{1994A&A...290..133V}
{van Paradijs}, J., \& {McClintock}, J.~E. 1994, \aap, 290, 133

\bibitem[{{Vaytet} {et~al.}(2011){Vaytet}, {O'Brien}, {Page}, {Bode}, {Lloyd},
  \& {Beardmore}}]{2011ApJ...740....5V}
{Vaytet}, N.~M.~H., {O'Brien}, T.~J., {Page}, K.~L., {et~al.} 2011, \apj, 740,
  5

\bibitem[{{Vrtilek} {et~al.}(1990){Vrtilek}, {Raymond}, {Garcia}, {Verbunt},
  {Hasinger}, \& {Kurster}}]{1990A&A...235..162V}
{Vrtilek}, S.~D., {Raymond}, J.~C., {Garcia}, M.~R., {et~al.} 1990, \aap, 235,
  162

\bibitem[{{Wade} \& {Hubeny}(1998)}]{1998ApJ...509..350W}
{Wade}, R.~A., \& {Hubeny}, I. 1998, \apj, 509, 350

\bibitem[{{White} {et~al.}(1995){White}, {Giommi}, {Heise}, {Angelini}, \&
  {Fantasia}}]{1995ApJ...445L.125W}
{White}, N.~E., {Giommi}, P., {Heise}, J., {Angelini}, L., \& {Fantasia}, S.
  1995, \apjl, 445, L125

\bibitem[{{Williams} {et~al.}(2004){Williams}, {Garcia}, {Kong}, {Primini},
  {King}, {Di Stefano}, \& {Murray}}]{2004ApJ...609..735W}
{Williams}, B.~F., {Garcia}, M.~R., {Kong}, A.~K.~H., {et~al.} 2004, \apj, 609,
  735

\bibitem[{{Williams} {et~al.}(2014{\natexlab{a}}){Williams}, {Lang},
  {Dalcanton}, {Dolphin}, {Weisz}, {Bell}, {Bianchi}, {Byler}, {Gilbert},
  {Girardi}, {Gordon}, {Gregersen}, {Johnson}, {Kalirai}, {Lauer}, {Monachesi},
  {Rosenfield}, {Seth}, \& {Skillman}}]{2014ApJS..215....9W}
{Williams}, B.~F., {Lang}, D., {Dalcanton}, J.~J., {et~al.} 2014{\natexlab{a}},
  \apjs, 215, 9

\bibitem[{{Williams} {et~al.}(2014{\natexlab{b}}){Williams}, {Darnley}, {Bode},
  {Keen}, \& {Shafter}}]{2014ApJS..213...10W}
{Williams}, S.~C., {Darnley}, M.~J., {Bode}, M.~F., {Keen}, A., \& {Shafter},
  A.~W. 2014{\natexlab{b}}, \apjs, 213, 10

\bibitem[{{Worters} {et~al.}(2007){Worters}, {Eyres}, {Bromage}, \&
  {Osborne}}]{2007MNRAS.379.1557W}
{Worters}, H.~L., {Eyres}, S.~P.~S., {Bromage}, G.~E., \& {Osborne}, J.~P.
  2007, \mnras, 379, 1557

\bibitem[{{Woudt} \& {Ribeiro}(2014)}]{2014ASPC..490.....W}
{Woudt}, P.~A., \& {Ribeiro}, V.~A.~R.~M., eds. 2014, Astronomical Society of
  the Pacific Conference Series, Vol. 490, {Stella Novae: Past and Future
  Decades} (San Francisco: Astronomical Society of the Pacific)

\end{thebibliography}

\end{document}